\documentstyle[12pt]{JHEP3}
\newcommand{\bej}[1]{ \begin{equation}\label{#1} }
\newcommand{\eej}{\end{equation}}
\newcommand{\beaj}[1]{\begin{eqnarray}\label{#1} }
\newcommand{\eeaj}{\end{eqnarray}}

\def\ZZZ{{\hskip-3pt\hbox{ Z\kern-1.6mm Z}}}
\def\zzz{{\hskip-3pt\hbox{ z\kern-1mm z}}}

\newcommand{\ra}{\rangle}
\newcommand{\la}{\langle}

\newcommand{\BB}{{\cal B}}

\newcommand{\AAA}{{\cal A}}

\newcommand{\MM}{{\cal M}}

\newcommand{\DD}{{\cal D}}

\newcommand{\OO}{{\cal O}}

\newcommand{\RR}{{\cal R}}
\newcommand{\NN}{{\cal N}}

\newcommand{\SSS}{{\cal S}}

\newcommand{\be}{\begin{equation}}
\newcommand{\ee}{\end{equation}}
\newcommand{\ben}{\begin{eqnarray}\displaystyle}
\newcommand{\een}{\end{eqnarray}}
\newcommand{\refb}[1]{(\ref{#1})}
\newcommand{\p}{\partial}
\newcommand{\sectiono}[1]{\section{#1}\setcounter{equation}{0}}

\def\one{{\hbox{ 1\kern-.8mm l}}}
\def\zero{{\hbox{ 0\kern-1.5mm 0}}}

\title{Boundary S-matrix in a $(2,0)$ theory of $AdS_{3}$ Supergravity}
\author{Payal Kaura $^{\clubsuit}$ and Bindusar Sahoo$^{\spadesuit}$\\
$^\clubsuit$Indian Institute Of Technology,\\
$\;$Roorkee 247667, India.\\
$\;$\email{pa123dph@iitr.ernet.in, payal.kaura@gmail.com}\\
$^\spadesuit$Harish-Chandra Research Institute, \\
$\;$Chhatnag Road., Jhunsi, \\
$\;$Allahabad 211019, India.\\
$\;$\email{bindusar@hri.res.in, bindusar@gmail.com}
}

\abstract{
We will discuss two inequivalent generalizations of the standard $(2,0)$ 
supergravity action \cite{Izquierdo:1994jz} to include 
gravitational Chern-Simons term. One 
is in the first order formalism where we treat $\omega_{M}^{~~ab}$ as 
independent and the other is in the second order formalism where 
$\omega_{M}^{~~ab}$ is determined in terms of other fields via a standard 
constraint equation. The two theories 
have different equations of motion and the solutions to the 
equations of motion of the first 
order theory spans only a subset of those of the second order theory. 
We will be interested 
in computing the boundary S-matrix, describing correlation functions 
in a dual conformal field theory, in this 
sector and hence we use the equations of motion 
coming out of the first order theory. We restrict ourselves to the 
gauge+fermionic sector of the theory to compute the boundary S-matrix. We will
 also look at the effect of higher derivative terms on the boundary S-matrix 
obtained using the standard supergravity action.}

\begin{document}
\baselineskip 3.5ex
\sectiono{Introduction} \label{intro}
Three dimensional $AdS_3$ supergravity has always been an interesting area of 
study since the discovery of BTZ black holes \cite{Banados:1992wn}. 
It has also played an important role in string theory since BTZ arises as 
a factor in the near horizon geometry of certain class of black holes in 
string theory \cite{Strominger:1997eq}. The entropy of BTZ 
black holes has been computed in two derivative theory of gravity 
\cite{Brown:1986nw} as well as higher derivative theories of gravity 
\cite{Saida:1999ec,Kraus:2005vz,Kraus:2005zm,Solodukhin:2005ah,Sahoo:2006vz,
Kraus:2006wn,Tachikawa:2006sz}. The statistical analysis of the entropy 
\cite{Brown:1986nw,Saida:1999ec,Kraus:2005vz,Kraus:2005zm,Solodukhin:2005ah,Kraus:2006wn} exploits the asymptotic symmetry of BTZ black hole and AdS/CFT 
correspondence \cite{Maldacena:1997re} 
whereas the gravity analysis in \cite{Sahoo:2006vz} and 
\cite{Tachikawa:2006sz} applies Wald's formula in 
the presence of gravitational Chern-Simons term. The entropy of BTZ black holes
 in the presence of arbitrary higher derivative terms has also been computed in
 \cite{Park:2006gt} from the first law of thermodynamics which also 
includes various aspects of statistical entropy.
All the above analysis reproduce the same result for the entropy of BTZ 
black holes which has remarkable similarity with the Cardy formula for the 
degeneracy of states in the two dimensional conformal field theory.

Later, Kraus and Larsen argued using AdS/CFT correspondence that if the theory 
has extended supersymmetry then the entropy of such black holes is given 
completely in terms of the coefficient of the gravitational Chern-Simons term
 and the coefficient of the Chern-Simons term involving the R-Symmetry gauge 
field \cite{Kraus:2005vz,Kraus:2005zm}. 
A bulk interpretation of their result was understood in 
\cite{David:2007ak} for 
theories with (0,4) supersymmetry and higher. The main conclusion of 
\cite{David:2007ak} 
was that the boundary S-matrix does not get renormalized by the addition of
 higher derivative terms. Since the boundary S-matrix are the only perturbative
 observables in the bulk theory, their result implied that the bulk action 
even does not get renormalized in the bulk. This in turn implies the 
non-renormalization of the black hole entropy. 

Later, in \cite{Gupta:2007th} Gupta and Sen came up with a result where the 
non-renormalization of the bulk action does not resort to the boundary 
S-matrix. They found that one can obtain a direct field redefinition in the 
bulk to absorb the higher derivative piece to bring the action to the standard
 form. The standard form of the action contains the Einstein-Hilbert piece 
with a negative cosmological constant term, gravitational Chern-Simons term 
and its supersymmetrization, Chern-Simons like term involving the gravitino. 
The standard form also has a 
Chern-Simons term involving the R-Symmetry 
gauge field if the theory 
has an extended supersymmetry. Such field redefinition does not affect the 
Chern-Simons terms but {\it {a priori}} it allows for the 
cosmological constant to 
change\footnote{For a theory of pure gravity, the effect of higher derivative 
terms will be a renormalization in the bare parameters like cosmological 
constant. This phenomena has also been observed in 
\cite{Park:2006pb,Park:2006zw} but this situation changes drastically 
in a theory of extended supergravity because the higher derivative terms 
does not affect the coefficients of Chern-Simons 
terms and since the cosmological constant is related to the coefficient of 
these Chern-Simons terms, even the cosmological constant is not modified 
in a theory of extended supergravity with arbitrary higher derivative terms.}. 
But the cosmological constant is determined in terms of the 
coefficient of the gravitational and gauge Chern-Simons term in a theory with 
extended supersymmetry. Hence, even 
the cosmological constant does not change 
by the field redefinitions.

Since the results of ``Kraus and Larsen'' and ``Gupta and Sen'' 
holds for all theories of
gravity with extended supersymmetry, we would expect that the boundary 
S-matrix does not get renormalized 
even for theories with lesser extended supersymmetry ({\it{e.g.}} 
theories 
with $(2,0)$ supersymmetry). The meaning of S-matrix not getting renormalized 
is however upto unitary transformation. Two S-matrices related by unitary 
transformations are the same and are related by redefinition of currents 
in the boundary. The field redefinition in the bulk induces this redefinition 
of currents in the boundary and hence the S-matrices 
calculated from two actions related by field redefinition in the bulk are the 
same upto unitary transformation. And since by the 
results of ``Gupta and Sen'' \cite{Gupta:2007th} restricted to $(2,0)$ theory,
 all $(2,0)$ actions are related by field redefinitions, this implies that 
the S-matrix with and without higher derivative terms are related by unitary 
transformations. This also implies the non-renormalization of central charges 
in the boundary CFT even for $(2,0)$ theories which in turn implies Kraus and 
Larsen's non-renormalization of black hole entropy even for $(2,0)$ theories.

However, one might wonder that what sort of higher derivative terms do 
not alter the boundary S-matrix computed using the standard supergravity 
action\footnote{We shall sometimes refer to the 
boundary S-matrix computed using the standard supergravity 
action as the standard boundary S-matrix.} and what sort of higher derivative terms alter the standard 
boundary S-matrix upto unitary transformation. We shall be interested in 
addressing this issue in our paper.

For boundary S-matrix in a $(0,4)$ theory of supergravity involving $SU(2)$ 
current correlators, this unitary transformation is exactly identity as 
observed in \cite{David:2007ak}. This observation however fails for a certain 
class of higher derivative terms in $(2,0)$ case. In order to distinguish such 
terms let us try to recap the results of \cite{David:2007ak}.

A theory of supergravity with $(0,4)$ supersymmetry has a $SU(2)$ R-symmetry 
and a $SU(2)$ gauge field $A_{M}^{a}$ corresponding to that. This gauge 
field sources the $SU(2)$ R-symmetry current in the boundary CFT. One then 
works with the standard Chern-Simons action $S_0$ and truncates to just the 
bosonic part. The gauge field equation of motion in this case becomes 
$F^{a}_{MN}=0$. In order to 
obtain the correlation function involving this current, one 
first obtains a 
solution to the gauge field equation of motion $F^{a}_{MN}=0$ with a 
boundary condition specified on $A_{z}^{a}$. After putting this 
solution in the action
 $S_0$ one obtains a functional $I[A_{z}^{(0)a}]$ of the boundary values 
$A_{z}^{(0)a}$. Then according to AdS/CFT correspondence 
$e^{-I[A_{z}^{(0)a}]}$ is interpreted as 
the partition function for calculating the correlation functions in the 
boundary \cite{Witten:1998qj,Gubser:1998bc} .

One can then add higher derivative gauge invariant term to the action $S_0$ as
\be \label{fullaction}
S=S_{0}+\lambda\int{d^{3}x F^{a}_{MN}K^{aMN}},
\ee
where $F^{a}_{MN}$ is the gauge covariant field strength. $K^{aMN}$ is some 
arbitrary term constructed out of field strengths, Riemann tensor etc. It 
is gauge covariant and hence it carries the gauge index $a$ which is required 
to make the full action gauge invariant. Thus $K^{aMN}$ should contain at 
least one power of $F^{a}_{MN}$. This implies that 
the additional terms in the action 
\refb{fullaction} contains at least two powers of $F^{a}_{MN}$ as a result 
of which it does not alter the equation of motion $F_{MN}^{a}=0$ 
which is also the equation of 
motion without the additional term. As a consequence of this the additional 
term vanishes on-shell and hence does not alter the boundary S-matrix.

The crux of the above observation is that additional higher derivative terms 
do not change the original equations of motion and hence vanish on-shell. 
But on the contrary for $(2,0)$ theories, as we will see later, there can be 
two sets of higher derivative terms
\begin{enumerate}
\item
Terms constructed out of field strengths covariant with respect to 
supersymmetry transformations under which the standard action remains invariant
\footnote{We will give an explicit definition and derivation of field strengths
 covariant with respect to the original supersymmetry transformation rules in 
section \ref{covfield}.}. Such terms will be the basic building blocks for 
constructing higher derivative terms which renders the full action 
to be invariant under the original 
supersymmetry transformation laws. We will see in the beginning of section 
\ref{highder} that such terms do not change 
the original equations of motion, and as a consequence of this, 
vanish on-shell and hence does not alter the boundary S-matrix. The 
unitary transformation, discussed earlier, in this case will be exactly 
identity.
\item
However, there can be other higher derivative terms which can render the full 
action to be invariant under a set of supersymmetry transformation laws 
different from the original supersymmetry transformation laws\footnote{We will sometimes refer to the original supersymmetry transformation laws as the 
``standard supersymmetry transformation laws''.}. We will see 
towards the end of section \ref{highder} that such terms can potentially 
change the equations of motion of the standard supergravity action and hence 
do not vanish on-shell, and as a result of this, can alter the boundary 
S-matrix calculated from the standard action. 
However, such terms can be removed by a redefinition 
of the fields\footnote{Field redefinition does not change the $AdA$ and 
$\bar{\psi}A\psi$ term but it can potentially change the $\bar{\psi} d\psi$ 
term. However $U(1)$ gauge invariance fixes the relative coefficient of 
the $\bar{\psi} d\psi$ and $\bar{\psi}A\psi$ term and hence even this 
coefficient is also not changed.} and hence, as discussed before, 
the change in S-matrix will be 
a non-trivial unitary transformation.
\end{enumerate}  
For the sake of completeness, we will compute, in the end, the 
boundary S-matrix involving the correlation 
function of the boundary currents from the standard action $S_0$. 
But in this case we cannot just restrict ourselves to the gauge sector. This is
 because the gauge Chern-Simons action for a $U(1)$ gauge field is 
\be \label{gauge}
S_{\texttt{gauge}}= -\frac{a_{L}}{2}\int d^3 x ~\epsilon^{MNP}A_{M}\p_{N}A_{P}
\ee
In this case we can arbitrarily scale $A_M$ to change the coefficient before 
it. This was not the case for $(0,4)$ theory in \cite{David:2007ak} because apart from the 
$A\wedge dA$ term in the action there was a $A\wedge A \wedge A$ term which 
forbids the arbitrary scaling of coefficient before the gauge Chern-Simons 
term. So, for $(2,0)$ theory we need to have in the action a term which 
contains a power of $A$ different from two. And indeed there is such a term 
in the action which is the coupling of gravitino with the gauge field. Thus 
we need to work with gauge and fermionic sector simultaneously and calculate 
correlation function involving the R-symmetry current $J(z)$ and supersymmetry 
currents $G^{(+)}(z),G^{(-)}(z)$ from the standard action $S_0$. 

We organize the paper as follows:
\begin{enumerate}
\item
In section \ref{action} we will discuss the generalization of standard 
$\NN=(2,0)$ supergravity action \cite{Izquierdo:1994jz} to include 
gravitational Chern-Simons term. We will see that we can have two 
inequivalent generalizations and we will discuss in great detail the 
differences between them.
\item
In section \ref{covfield} we will define and obtain the field 
strengths and Riemann tensor covariant with respect to standard supersymmetry 
transformation laws. These will form the building blocks of 
constructing supersymmetric invariant higher derivative terms respecting the 
standard supersymmetry transformation laws. 
\item
In section \ref{highder} we discuss the implication of addition of 
higher derivative terms respecting supersymmetry on the boundary S-matrix 
which shall arise from the standard supergravity action. 
We will discuss two sets of higher 
derivative terms. First we will consider terms which 
are constructed out of the super-covariant field 
strengths discussed in section \ref{covfield} and we will see that they 
do not alter the standard boundary S-matrix. In the end we will consider more 
general terms which alter the standard 
boundary S-matrix by a non-trivial unitary 
transformation.
\item
To conclude, we will compute the boundary S-matrix describing correlation 
functions involving R-symmetry current $J(z)$ and supersymmetry current 
$G^{(+)}(z)$, $G^{(-)}(z)$ from the 
standard supergravity action in section \ref{correl}. 
In particular, we calculate the correlators 
$\left\la J(z)J(w)\right\ra$, $\left\la G^{+}(z)G^{-}(w)\right\ra$ and 
$\left\la J(z)G^{+}(v)G^{-}(w) \right\ra$. We will see that the results match
 with the conformal field theory results.
\end{enumerate}

\sectiono{$\NN=(2,0)$ supergravity action} 
\label{action}
We shall be working with a generalization of standard $(2,0)$ supergravity 
\cite{Izquierdo:1994jz} to include gravitational Chern-Simons term
\footnote{Such a theory is 
also known as the cosmological topological massive supergravity because of the 
presence of a massive excitation.}. Such a generalization has been obtained in 
\cite{Deser:1982sv} for the simple case of $\NN=1$ supersymmetry. We shall be 
interested in the case having $(2,0)$ supersymmetry. 

Before trying to obtain a $(2,0)$ supersymmetric generalization of 
topologically massive supergravity with a cosmological constant---which we shall 
refer to as cosmological topologically massive supergravity, 
let us first look at the 
standard $(2,0)$ supergravity without gravitational Chern-Simons term 
\cite{Izquierdo:1994jz}\footnote{We will often refer to this as the cosmological topological massless supergravity because of the absence of a massive excitation.}. The field contents are the vierbeins $e_{M}^{~~a}$, 
$U(1)$ gauge field $A_{\mu}$, and a complex Rarita-Schwinger field $\psi_{M}$. 
The action written in terms of 
the fields $e_{M}^{a}$, $\psi_{M}$, $\bar{\psi_{M}}$, $A_{\mu}$ is
\be \label{e2.1}
S_{0}=\int d^3 x \left[eR+2m^{2}e-\frac{1}{2m}\epsilon^{MNP}A_{M}\p_{N}A_{P}+
\frac{i}{4m} \epsilon^{MNP}\left(\bar{\psi}_{M}(\DD_{N}\psi_{P})-
(\DD_{N}\bar{\psi}_{M})\psi_{P}\right)\right] \nonumber \\
\ee
We follow the following convention for Riemann tensor, Ricci tensor and 
Ricci scalar
\ben \label{Ricci}
&& R_{MN}^{~~~~ab}\equiv 2\partial_{[M}\omega_{N]}^{~~ab}+2\omega_{[M}^{~~ac}\omega_{N]}^{~~db}\eta_{cd}, \nonumber \\
&& R_{M}^{~~a}\equiv e^{N}_{~~b}R_{MN}^{~~~~ab}, \nonumber \\
&& R \equiv e^{M}_{~~a}e^{N}_{~~b}R_{MN}^{~~~~ab},
\een
$\omega_{M}^{~~ab}$ is the spin-connection and is determined in terms 
of $e_{M}^{~~a}$ and $\psi_{M}$ from the equation
\be \label{spinconnection}
\partial_{[N}e_{P]}^{~~a}+\omega_{[N}^{~~ab}e_{P]b}=-\frac{i}{8m}\bar{\psi}_{[N}\gamma^{a}\psi_{P]}
\ee
and $e$ is the determinant of $e_{M}^{~~a}$ given by \footnote{We denote the 
curved space index $M$ by $0,1,2$ and flat space index $a$ by 
$\hat{0},\hat{1},\hat{2}$. The epsilon symbol with curved space index is 
denoted by $\epsilon$ and the epsilon symbol involving flat index is 
denoted by $\varepsilon$. The index in both cases is raised and lowered using 
$\eta_{MN}$ and $\eta_{ab}$ and we follow the mostly positive convention 
i.e $\eta=diag(-1,1,1)$.}
\be \label{edef}
e=\det\left(e_{M}^{~~a}\right)=\frac{1}{6}\epsilon^{MNP}\varepsilon_{abc}e_{M}^{~~a}e_{N}^{~~b}e_{P}^{~~c}, \quad \epsilon^{012}=1 \quad 
\varepsilon_{\hat{0}\hat{1}\hat{2}}=1
\ee
$\DD_{M}\psi_{N}$ and $\DD_{M}\bar{\psi}_{N}$ are defined as
\ben \label{e2.2}
\DD_{M}\psi_{N} &=& \p_{M}\psi_{N}-\frac{1}{2}B_{M}^{~~a}\gamma_{a}\psi_{N}+
\frac{i}{2}A_{M}\psi_{N}, \nonumber \\
\DD_{M}\bar{\psi}_{N} &=& \p_{M}\bar{\psi}_{N}+
\frac{1}{2}B_{M}^{~~a}\bar{\psi}_{N}\gamma_{a}-
\frac{i}{2}A_{M}\bar{\psi}_{N},
\een
$B_{M}^{~~a}$ is given by
\be \label{e2.3}
B_{M}^{~~a}=\frac{1}{2}\varepsilon^{abc}\omega_{Mbc}-m e_{M}^{~~a}
\ee
$\gamma_{a}$ satisfies the algebra
\be \label{gamma}
\left \{\gamma_{a},\gamma_{b} \right \}=2 \eta_{ab} \quad 
\left[\gamma_{a},\gamma_{b} \right]=-2\varepsilon_{abc}\gamma^{c}
\ee
We can write the gravity part of the action \refb{e2.1} 
in a pure Chern-Simons form 
by defining the gauge field $B_{M}^{\prime~~a}$ in addition to $B_{M}^{~~a}$ 
as:
\be \label{e2.4}
B_{M}^{\prime~~a}=\frac{1}{2}\varepsilon^{abc}\omega_{Mbc}+m e_{M}^{~~a}
\ee
Then the action \refb{e2.1} takes the following form
\footnote{One can check by expanding $B_{M}^{\prime~~a}$ and $B_{M}^{~~a}$ in 
the action \refb{e2.5} in terms of $\omega_{M}^{~ab}$ and $e_{M}^{~a}$ that 
one indeed gets back the action \refb{e2.1}.} 
\cite{Fjelstad:2001je,Witten:1988hc,Witten:2007kt}:
\ben \label{e2.5}
S_0 &=& -\frac{1}{m}\int d^3 x ~{\epsilon^{MNP}\left[\frac{1}{2}B_{M}^{~~a}\p_{N}B_{P}^{~~b}
\eta_{ab}+\frac{1}{6}\varepsilon_{abc}B_{M}^{~~a}B_{N}^{~~b}B_{P}^{~~c}\right]} \nonumber \\
&& +\frac{1}{m}\int d^3 x ~{\epsilon^{MNP}\left[\frac{1}{2}B_{M}^{\prime~~a}\p_{N}B_{P}^{\prime~~b}\eta_{ab}+\frac{1}{6}\varepsilon_{abc}B_{M}^{\prime~~a}B_{N}^{\prime~~b}B_{P}^{\prime~~c}\right]} \nonumber \\
&& -\frac{1}{2m}\int d^3 x ~{\epsilon^{MNP}A_{M}\p_{N}A_{P}} \nonumber \\
&& +\frac{i}{4m}\int d^3 x ~{\epsilon^{MNP}
\left(\bar{\psi}_{M}(\DD_{N}\psi_{P})-(\DD_{N}\bar{\psi}_{M})\psi_{P}\right)}
\een
Note that in the above action, $B_{M}^{~~a}$ and $B_{M}^{\prime~~a}$ are 
not independent. Hence, this form of the action is not very useful. But we will
 see later that we can have a different formulation where we treat 
$\omega_{M}^{~~ab}$ as an independent field and hence 
$B_{M}^{~~a}$ and $B_{M}^{\prime~~a}$ becomes independent. In that case, the 
standard supergravity action in the above Chern-Simons form \refb{e2.5} 
is useful. However, \refb{e2.5}
 form of the action in the present formulation is useful in some context 
like looking at the supersymmetry invariance etc. But whenever we use the above
 form of the action in the present formulation, we should always keep in mind 
that $B_{M}^{~~a}$ and $B_{M}^{\prime~~a}$ are 
not independent. 

The action \refb{e2.1}, \refb{e2.5} is invariant under the supersymmetry 
transformations
\ben \label{e2.51}
\delta_{\epsilon}^{(Q)}e_{M}^{~a} &=& -\frac{i}{8m}\left(\bar{\epsilon}\gamma^{a}\psi_{P}-\bar{\psi_{P}}\gamma^{a}\epsilon\right) \nonumber \\
\delta_{\epsilon}^{(Q)}A_{P} &=& \frac{1}{4}\left(\bar{\epsilon}\psi_{P}-\bar{\psi_{P}}\epsilon\right), \nonumber \\
\delta_{\epsilon}^{(Q)}\psi_{P} &=& \DD_{P}\epsilon, \nonumber \\
\delta_{\epsilon}^{(Q)}\bar{\psi}_{P} &=& \DD_{P}\bar{\epsilon}, \nonumber \\
\een
Using \refb{spinconnection} one can deduce the following supersymmetry 
transformation on the dependent gauge field $\omega_{M}^{~~ab}$
\ben \label{e2.52}
\delta_{\epsilon}^{(Q)}\omega_{P}^{~~ab} &=&-\frac{i}{8}\varepsilon^{abc}\left(\bar{\epsilon}\gamma_{c}\psi_{P}-\bar{\psi_{P}}\gamma_{c}\epsilon\right) \nonumber \\
&& +\frac{i}{16m}\left[\bar{\epsilon}\gamma_{P}G^{ab}
-\bar{G}^{ab}\gamma_{P}\epsilon\right] \nonumber \\
&& +\frac{i}{8m}\left[\bar{\epsilon}\gamma^{[a}G_{P}^{~~b]}-
\bar{G}_{P}^{~~[b}\gamma^{a]}\epsilon\right] \nonumber \\
&\equiv& -\frac{i}{8}\varepsilon^{abc}\left(\bar{\epsilon}\gamma_{c}\psi_{P}-\bar{\psi_{P}}\gamma_{c}\epsilon\right)+ C_{P}^{~~ab}(\epsilon,G)
\een
Where $G_{MN}$ is the field strength associated with the Rarita-Schwinger 
field $\psi_M$ defined in \refb{EOM}. The last line in the above equation 
defines $C_{P}^{~~ab}(\epsilon,G)$. The supersymmetry transformations of 
the fields $B_{M}^{~~a}$ and 
$B_{M}^{\prime~~a}$ defined in (\ref{e2.3}, \ref{e2.4}) takes the form
\ben \label{e2.53}
\delta_{\epsilon}^{(Q)}B_{P}^{~~a} &=& \frac{i}{4}\left(\bar{\epsilon}\gamma^{a}\psi_{P}-\bar{\psi_{P}}\gamma^{a}\epsilon\right)
+\frac{1}{2}\varepsilon^{abc}C_{Pbc}, \nonumber \\
\delta_{\epsilon}^{(Q)}B_{P}^{\prime~~a} &=& \frac{1}{2}\varepsilon^{abc}C_{Pbc}
\een
The equation of motion derived from the action \refb{e2.1} is
\ben \label{EOM}
&& \RR_{a}^{~~M}-\frac{1}{2}\RR e_{a}^{~~M}-m^{2}e_{a}^{~~M} = 0, \nonumber \\
&& \widehat{F}_{MN}\equiv 2\p_{[M}A_{N]}-\frac{1}{2}\bar{\psi}_{[M}\psi_{N]}=0,
 \nonumber \\
&& G_{MN}\equiv 2\DD_{[M}\psi_{N]}=0 \quad 
\bar{G}_{MN}\equiv 2\DD_{[M}\bar{\psi}_{N]}=0
\een
where
\ben \label{eom1}
&& \RR_{MN}^{~~~~ab}= R_{MN}^{~~~~ab}+\frac{i}{4}\varepsilon^{abc}\bar{\psi}_{[M}\gamma_{c}\psi_{P]}, \nonumber \\
&& \RR_{M}^{~~a}=e_{b}^{~~N}\RR_{MN}^{~~~~ab}, \nonumber \\
&& \RR = e_{a}^{~~M}\RR_{M}^{~~a},
\een
$\widehat{F}_{MN}$ and $G_{MN}$ as we will see later are supersymmetric 
covariant field strengths for the gauge field $A_M$ and Rarita
-Scwhinger field $\psi_M$ respectively and $\RR_{MN}^{~~~~ab}$ is the 
supersymmetric covariant Riemann tensor modulo some terms proportional to 
$G_{MN}$. It is straightforward to check from \refb{EOM} that $\RR_{NP}$ 
satisfies
\be \label{eom1.2}
D_{M}\RR_{NP}=0
\ee
Where $D_{M}$ is the usual covariant derivative defined using the 
torsion free connections. The field equation corresponding to 
$\RR_{MN}^{~~~~ab}$ can also be written in a more convenient way as
\ben \label{eom2}
&& \widehat{\RR}_{M}^{~~a}=0 \nonumber \\
\texttt{Where} && \widehat{\RR}_{MN}^{~~~~ab}=\RR_{MN}^{~~~~ab}+
2m^2 e_{[M}^{~~a}e_{N]}^{~~b}
\een
In the above analysis of the theory of supergravity without the gravitational 
Chern-Simons 
term, we treated $\omega_{M}^{~~ab}$ as defined through the equation 
\refb{spinconnection} and the only independent fields in the theory were 
$e_{M}^{~~a}$, $\psi_M$ and $A_M$. This formulation is known in the literature
 as the ``second order formulation'' since it gives second order equations 
in the gravity sector. We will also resort to the supersymmetry 
transformations \refb{e2.52} and 
\refb{e2.53} for $\omega_{M}^{~~ab}$, $B_{M}^{~~a}$ and $B_{M}^{\prime~~a}$ as 
the ``second order supersymmetry transformations''. 

In a different formulation we can 
treat $\omega_{M}^{~~ab}$ as an independent field in \refb{e2.1}.
 The variation of the action with respect to this field will give rise to the 
constraint equation \refb{spinconnection} and the variation with respect 
to other fields will give rise to the other equations in \refb{EOM}. All these 
equations are first order since we are treating $\omega_{M}^{~~ab}$ as 
independent. This formulation is known as the ``first order formulation''. In 
this formulation $B_{M}^{~~a}$ and $B_{M}^{\prime~~a}$ are independent and, as
 discussed before, 
the Chern-Simons form of the action \refb{e2.5} is a useful form to use.

We now discuss the supersymmetry invariance of the action under 
supersymmetry transformation independently 
defined for $\omega_{M}^{~~ab}$ 
\be \label{e2.531}
\delta_{\epsilon}^{(Q)}\omega_{P}^{~~ab}=-\frac{i}{8}\varepsilon^{abc}\left(\bar{\epsilon}\gamma_{c}\psi_{P}-\bar{\psi_{P}}\gamma_{c}\epsilon\right)
\ee
This also implies the following supersymmetry transformation for the 
fields $B_{m}^{~~a}$ and $B_{M}^{\prime~~a}$
\ben \label{e2.5311}
\delta_{\epsilon}^{(Q)}B_{P}^{~~a} &=& \frac{i}{4}\left(\bar{\epsilon}\gamma^{a}\psi_{P}-\bar{\psi_{P}}\gamma^{a}\epsilon\right)
\nonumber \\
\delta_{\epsilon}^{(Q)}B_{P}^{\prime~~a} &=& 0
\een
It is quite easy to check that the action \refb{e2.1}, \refb{e2.5} is 
invariant 
with respect to 
this transformation along with \refb{e2.51} 
when we treat $\omega_{M}^{~~ab}$ as an 
independent gauge field. We will refer to the 
supersymmetry transformations \refb{e2.531} and \refb{e2.5311} as the 
``first order supersymmetry transformations''. 
From \refb{e2.5311} we see that the supersymmetry 
transformation of $B_{M}^{\prime~~a}$ vanishes and this can be interpreted as 
belonging to the right sector of the theory where we do not have any 
supersymmetry whereas the rest of the fields ($B_{M}^{~~a}$, $\psi_M$, $A_M$)
 belongs to the left sector which has $\NN=2$ supersymmetry. This justifies 
the $(2,0)$ nature of the theory.

This theory is supersymmetric in both the formulations. 
Both the formulations also give rise to the same equations of motion. 
In particular, the 
$\omega_{M}^{~~ab}$ equation of motion in the first order formulation is 
the same as the equation that determines $\omega_{M}^{~~ab}$ in terms of 
$e_{M}^{~~a}$ and $\psi_M$ in the second order formulation so 
that we can still interprete $\omega_{M}^{~~ab}$ as the spin connection. 

Let us now try to make appropriate changes in \refb{e2.5} to 
obtain the gravitational Chern-Simons term and its 
supersymmetric generalization. Following \cite{Witten:2007kt,David:2007ak}, 
we make the following changes
\ben \label{TMG}
\SSS_{0} &=& -a_L\int d^3 x ~{\epsilon^{MNP}\left[\frac{1}{2}B_{M}^{~~a}\p_{N}B_{P}^{~~b}
\eta_{ab}+\frac{1}{6}\varepsilon_{abc}B_{M}^{~~a}B_{N}^{~~b}B_{P}^{~~c}\right]} \nonumber \\
&& +a_R\int d^3 x ~{\epsilon^{MNP}\left[\frac{1}{2}B_{M}^{\prime~~a}\p_{N}B_{P}^{\prime~~b}\eta_{ab}+\frac{1}{6}\varepsilon_{abc}B_{M}^{\prime~~a}B_{N}^{\prime~~b}B_{P}^{\prime~~c}\right]} \nonumber \\
&& -\frac{a_L}{2}\int d^3 x ~{\epsilon^{MNP}A_{M}\p_{N}A_{P}} \nonumber \\
&& +\frac{ia_L}{4}\int d^3 x ~{\epsilon^{MNP}
\left(\bar{\psi}_{M}(\DD_{N}\psi_{P})-(\DD_{N}\bar{\psi}_{M})\psi_{P}\right)}
\nonumber \\
&& a_L=K+\frac{1}{m} \quad \quad a_R=-K+\frac{1}{m}
\een
After replacing $B_{M}^{~~a}$ and $B_{M}^{\prime~~a}$ from \refb{e2.3} and 
\refb{e2.4}
 one can see that one indeed gets the gravitational Chern-Simons term
\ben \label{TMG0.1}
&& \SSS_{0}=\int d^3 x \left[eR+2m^{2}e-\frac{a_{L}}{2}\epsilon^{MNP}A_{M}\p_{N}A_{P}+
\frac{i}{4} a_{L}\epsilon^{MNP}\left(\bar{\psi}_{M}(\DD_{N}\psi_{P})-
(\DD_{N}\bar{\psi}_{M})\psi_{P}\right)\right] \nonumber \\
&& -K\int d^3 x ~\epsilon^{MNP}\left[\left(\frac{1}{2}\omega_{Mcd}\p_{N}\omega_{P}^{~~dc}+\frac{1}{3}\omega_{Mbc}\omega_{N}^{~~cd}\omega_{Pd}^{~~b}\right)
+m^2 e_{M}^{~~a}\left(\p_N e_{P}^{~~a}+ \omega_{Nac}e_{P}^{~~c}\right)\right], 
\nonumber \\
&& a_{L}=K+\frac{1}{m},
\een
In the first order formalism, when $\omega_{M}^{~~ab}$ is treated as an 
independent field, the first order supersymmetry transformation 
\refb{e2.5311} and \refb{e2.51} of the left sector which 
comprises of $B_{M}^{~~a}$, $\psi_M$ and $A_M$ do not talk with the 
right sector comprising of $B_{M}^{\prime~~a}$. In the above action we have 
just changed the relative coefficients between the left and right sector 
fields without changing the relative coefficients between the fields of a 
particular sector. We therefore expect the above action to be invariant under 
the ``first order'' supersymmetry variations \refb{e2.531}, \refb{e2.5311}, 
\refb{e2.51}. We find that the action \refb{TMG}, \refb{TMG0.1} is indeed 
invariant under the first order supersymmetry transformations 
\refb{e2.531}, \refb{e2.5311}, \refb{e2.51}. In order to obtain the 
equation of motion we first vary the 
action \refb{TMG0.1}
\ben \label{var}
\delta \SSS_{0} &=& 2\int d^3 x ~\epsilon^{MNP}\delta \omega_{M}^{~~c}
\left[\p_N e_{Pc}+\omega_{Nc}^{~~~d}e_{Pd}
+\frac{i}{8m}\bar{\psi}_{N}\gamma_{c}\psi_{P}\right] \nonumber \\
&& -K\int d^3 x ~\delta \omega_{M}^{~~c}~e\left[\RR e^{M}_{~~c}+2 m^2 e^{M}_{~~c}
-2\RR^{M}_{~~c}\right] \nonumber \\
&& +\int d^3 x 
~ \delta e_{M}^{~~a}~e\left[\RR e^{M}_{~~a}+2 m^2 e^{M}_{~~a}
-2\RR^{M}_{~~a}\right] \nonumber \\
&& -2Km^2\int d^3 x~ \delta e_{M}^{~~a} \epsilon^{MNP}
\left[\p_N e_{Pa}+\omega_{Na}^{~~~d}e_{Pd}
+\frac{i}{8m}\bar{\psi}_{N}\gamma_{a}\psi_{P}\right] \nonumber \\
&& -a_{L}\int d^3 x \delta A_M \epsilon^{MNP}
\left(\p_N A_P-\frac{1}{4}\bar{\psi}_{N}\psi_{P}\right)
+\frac{ia_L}{2}\int d^3 x 
\delta \bar{\psi}_{M} \epsilon^{MNP} \DD_{N}\psi_P \nonumber \\
&& +\frac{ia_L}{2}\int d^3 x ~\epsilon^{MNP} \DD_{N}\bar{\psi}_P \delta \psi_M
\een
In the first order formulation when we are treating 
$\omega_{M}^{~~ab}$ as an independent field, we should 
set all the variations independently to zero to get the equations of motion. 
We get
\ben \label{cheom}
&& 2\epsilon^{MNP}\left[\p_N e_{Pc}+\omega_{Nc}^{~~~d}e_{Pd}
+\frac{i}{8m}\bar{\psi}_{N}\gamma_{c}\psi_{P}\right]
-K e\left[\RR e^{M}_{~~c}+2 m^2 e^{M}_{~~c}
-2\RR^{M}_{~~c}\right]= 0 \nonumber \\
&& -2Km^2 \epsilon^{MNP}
\left[\p_N e_{Pa}+\omega_{Na}^{~~~d}e_{Pd}
+\frac{i}{8m}\bar{\psi}_{N}\gamma_{a}\psi_{P}\right]
+e\left[\RR e^{M}_{~~a}+2 m^2 e^{M}_{~~a}
-2\RR^{M}_{~~a}\right]=0 \nonumber \\
&& F_{MN}\equiv 2\p_{[M}A_{N]}=\frac{1}{2}\bar{\psi}_{[M}\psi_{N]}, 
\nonumber \\
&& G_{MN}\equiv 2\DD_{[M}\psi_{N]}=0
\een
When $Km\neq \pm 1$ we can take linear combinations of first two 
equations and we get
\ben \label{cheom1}
&& \partial_{[N}e_{P]}^{~~a}+\omega_{[N}^{~~ab}e_{p]b}=-\frac{i}{8m}\bar{\psi}_{[N}\gamma^{a}\psi_{P]}, \nonumber \\
&& \RR_{a}^{~~M}-\frac{1}{2}\RR e_{a}^{~~M}-m^{2}e_{a}^{~~M} = 0,
\een
We thus see that when $Km\neq \pm1$ we get the same equation that we got 
in our 
analysis of the supergravity theory without gravitational Chern-Simons term. 
However, when $Km=\pm 1$ the first two equations of \refb{cheom} are degenerate 
and instead of two 
there is one equation governing $e_{M}^{~~a}$ and $\omega_{M}^{~~ab}$. 
This can also be seen from the action \refb{TMG} written in terms of 
$B_{M}^{~~a}$ and $B_{M}^{\prime~~a}$. When $Km=\pm 1$ either $a_L$ or $a_R$
 vanishes. Then we have either $B_{M}^{~~a}$ or $B_{M}^{\prime~~a}$ present 
in the action describing the gravity sector. Thus the degree of freedom 
required for a theory of gravity is reduced and we cannot have a theory 
of gravity. Hence, 
the first order formulation fails for $Km=\pm 1$\footnote{Such a phenomena has
 also been observed in a three dimensional theory of pure gravity in 
\cite{Park:2006gt}.}. However, at a generic point 
not satisfying $Km=\pm 1$ we get a perfectly sensible first order theory which 
is supersymmteric and gives the equations of motion which coincides with 
the equation of motion derived for the theory of supergravity without 
gravitational Chern-Simons term.

We cannot add higher derivative terms to the supergravity action in the 
first order formulation. The equation 
for $\omega_{M}^{~~ab}$ will become dynamical 
and there cannot be simple algebraic dependence of $\omega_{M}^{~~ab}$ on 
$e_{M}^{~~a}$ and $\psi_M$ as in \refb{spinconnection} and hence we cannot 
interpret $\omega_{M}^{~~ab}$ as the spin connection. 
Addition of higher derivative terms will require 
us to go the second order picture.

In the second order formulation we work with $e_{M}^{~~a}$, 
$\omega_{M}^{~~ab}$,
 $\psi_M$ and $A_M$ with $\omega_{M}^{~~ab}$ determined in terms of the 
other fields through \refb{spinconnection}. Substituting 
$\omega_{M}^{~~ab}$ by the constraint equation \refb{spinconnection} in 
\refb{TMG0.1} gives rise to the action
\ben \label{TMG1}
&& \SSS_{0}=\int d^3 x \left[eR+2m^{2}e-\frac{a_{L}}{2}\epsilon^{MNP}A_{M}\p_{N}A_{P}+
\frac{i}{4} a_{L}\epsilon^{MNP}\left(\bar{\psi}_{M}(\DD_{N}\psi_{P})-
(\DD_{N}\bar{\psi}_{M})\psi_{P}\right)\right] \nonumber \\
&& -K\int d^3 x ~\epsilon^{MNP}\left[\left(\frac{1}{2}\omega_{Mcd}\p_{N}\omega_{P}^{~~dc}+\frac{1}{3}\omega_{Mbc}\omega_{N}^{~~cd}\omega_{Pd}^{~~b}\right)-
\frac{im}{8}e_M^{~~a}\bar{\psi}_N\gamma_{a}\psi_P\right], 
\nonumber \\
&& a_{L}=K+\frac{1}{m},
\een
However, the action \refb{TMG} or \refb{TMG1} is not invariant under the 
``second order supersymmetry transformations'' \refb{e2.51}, \refb{e2.52}, 
\refb{e2.53}.
Varying the action \refb{TMG} with respect to these 
transformations, we get
\be \label{TMG2}
\delta_{\epsilon}^{(Q)}\SSS_{0}= 
\frac{K}{2}\int~d^3 x ~\epsilon^{MNP}C_{Mbc}(\epsilon,G)
~\widehat{\RR}_{NP}^{~~~~bc}
\ee
This suggests that we should add to the action \refb{TMG} or \refb{TMG1}, 
terms constructed out of the field strengths 
($\widehat{F}_{MN}, G_{MN}, \widehat{\RR}_{NP}^{~~~~bc}$) such that the 
variation of such terms exactly cancel the variation of $\SSS_0$ obtained in 
\refb{TMG2}. We will not require what exactly these terms are, but 
we will certainly
 need the fact that such terms are constructed out of the above mentioned 
field strengths and not 
the gauge fields. So these will fall into the category of ``higher derivative 
terms'' whose effect we will see later. Thus the complete cosmological 
topologically massive action of gravity invariant 
under the supersymmetry transformations \refb{e2.51}-\refb{e2.53} is
\be \label{TMG3}
\SSS=\SSS_{0}+
\SSS_{1}\left[\widehat{F}_{MN}, G_{MN}, \widehat{\RR}_{NP}^{~~~~bc}\right]
\ee
Since $\SSS_{1}$ is higher in derivatives, we can use the results of 
\cite{Gupta:2007th} 
to argue that this term can be removed by an explicit field 
redefinition. In that case we suspect that $\SSS_{0}$ can still be 
supersymmetric even in the second order 
formulation by appropriately changing the transformation rules 
\refb{e2.51}-\refb{e2.53}. Since we have 
not been able to find the modified transformation rules, we would rather 
keep the existing transformation rules and allow for the presence of 
$\SSS_{1}$ in the full supergravity action.

In order to get the equations of motion constructed out of $\SSS_{0}$ 
in the second order 
formulation we can use the variation \refb{var}, but keeping in mind that 
$\delta \omega_{M}^{~~ab}$ is not independent and should be determined in 
terms of $\delta e_{M}^{~~a}$ and $\delta \psi_M$ from the constraint 
equation \refb{spinconnection}. After using the constraint 
\refb{spinconnection}, the variation \refb{var} takes the form
\ben \label{var2} 
\delta \SSS_{0} &=& -K\int d^3 x ~\delta \omega_{M}^{~~c}~e\left[\RR e^{M}_{~~c}+2 m^2 e^{M}_{~~c}
-2\RR^{M}_{~~c}\right] \nonumber \\
&& +\int d^3 x 
~ \delta e_{M}^{~~a}e\left[\RR e^{M}_{~~a}+2 m^2 e^{M}_{~~a}
-2\RR^{M}_{~~a}\right] \nonumber \\
&& -a_{L}\int d^3 x \delta A_M \epsilon^{MNP}
\left(\p_N A_P-\frac{1}{4}\bar{\psi}_{N}\psi_{P}\right)
+\frac{ia_L}{2}\int d^3 x 
\delta \bar{\psi}_{M} \epsilon^{MNP} \DD_{N}\psi_P \nonumber \\
&& +\frac{ia_L}{2}\int d^3 x ~\epsilon^{MNP} \DD_{N}\bar{\psi}_P \delta \psi_M
\een
To obtain the complete equation of motion 
systematically we should first express $\delta \omega_{M}^{~~ab}$ in terms 
of $\delta e_{M}^{~~a}$ and $\delta \psi_M$ and put it in the above variation. 
Then collect all the terms proportional to the variations of 
$e_{M}^{~~a}$, $\psi_{M}$ and 
$A_M$ and set them to zero. This will certainly not change the $A_M$ equations 
because $\delta \omega_{M}^{~~ab}$ does not contain $\delta A_M$ but it 
will certainly change the $e_{M}^{~~a}$ and $\psi_{M}$ equations of motion. 
The equation, in general will be 
quite complicated involving the Cotton tensor 
\cite{Cotton:1899co,Guralnik:2003we} and 
Cottino-vector spinor \cite{Gibbons:2008vi}. However, one can easily 
check that the 
equation of motion \refb{EOM} obtained for the topologically massless 
case (which is also the equation of motion of $\SSS_{0}$ in the 
first order formulation) still 
extremizes the variation of $\SSS_{0}$ in the second order formulation 
and hence gives a sector of solution 
for the theory with gravitational Chern-Simons term 
in the second order formulation. This is contrary to the case of 
supergravity without gravitational Chern-Simons term
where both the ``first order'' and 
``second order'' formulation give rise to the same equations of motion. 
In the presence of gravitational Chern-Simons term, we saw that 
in the ``first order'' formulation the equations of motion are the same 
as that of the theory without gravitational Chern-Simons term. 
Whereas, in the ``second order'' formulation the equations get modified and 
the earlier equations span only a part of the full spectrum\footnote{Since 
the equations of motion in the first order theory will be the same as 
that of the theory without gravitational Chern-Simons term, we will 
still call this theory as the cosmological topological massless supergravity. 
The nontrivial massive excitation arises when we go to the second order 
formulation of the theory with gravitational Chern-Simons term and we will
call this second order formulation as the cosmological topological 
massive supergravity.}. 

We see two major differences in the first order and second order 
formulation of a theory of $(2,0)$ supergravity with gravitational 
Chern-Simons term. One is in the supersymmetry invariance of $\SSS_{0}$. 
The other is in the equations of motion spanned by both of them. Such 
differences in both the formulations in a theory of topologically massive 
gravity (TMG) and topologically massive electrodynamics (TME), as far as 
equations of motion is concerned, 
has been previously observed in \cite{Deser:2006ht}. 
However, later in \cite{Kiriushcheva:2006df} it was argued that the 
definition of the ``first order theory'' in 
\cite{Deser:2006ht} is not the correct first order 
formulation of the ``second order theory'' because both of them give 
different equations of motion. The correct first order formulation will 
include several other non-dynamical fields. They found the additional fields 
required for the case of TME. Finding the additional fields 
required for the correct first order formulation of a theory of 
gravity in the presence of gravitational Chern-Simons term in 
general will be quite involved and as per our knowledge there is no such 
formulation yet.

However, we will keep our definition of first order and second order 
formalism without calling one as the equivalent of the other. As we have seen 
before, the supergravity spectrum spanned by our first order theory belongs 
to a subset of the second order theory.
We will be interested in the correlation functions of 
conformal field theory operators dual to this sector and 
hence we will use the equations of motion \refb{EOM} in our analysis of 
boundary S-matrix.


\sectiono{Supersymmetric covariant field strengths}\label{covfield}
In this section we will try to covariantize the various field strengths 
($R_{MN}^{~~ab}$, $F_{MN}$, $G_{MN}$, $\bar{G}_{MN}$) with respect to 
the standard second order supersymmetry 
transformations (\ref{e2.51}, \ref{e2.52}, \ref{e2.53}) considered in the 
previous section. First let us try to 
understand what do we mean by field strengths covariant with respect to 
supersymmetry. Normally the field strengths, {\it {for example}} Riemann 
tensor $R_{MN}^{~~~~ab}$ or gauge field strength 
${F}_{MN}\equiv 2\p_{[M}A_{N]}$, 
are not covariant with respect to supersymmetry i.e. under standard 
supersymmetry 
transformations (\ref{e2.51}, \ref{e2.52}, \ref{e2.53}) the above mentioned 
field strengths will give rise to terms proportional to partial 
derivatives of the supersymmetry transformation parameter. Thus we have to
 add certain terms to the above mentioned field strengths so that this 
non-covariant behavior gets canceled and we get perfectly covariant field 
strengths whose supersymmetry transformation gives rise to terms containing 
the supersymmetry transformation parameter and 
other supersymmetric covariant field strengths.
 As a result of this, if we add any term constructed out of these field 
strengths to the action, under supersymmetry it will give rise to terms 
involving the supersymmetry parameter and other super-covariant field 
strengths. This non-invariance under supersymmetry can be canceled by adding 
other terms to the action constructed out of the super-covariant field 
strengths such that its supersymmetry variation exactly cancels the 
supersymmetry variation of the term that we initially added to the action. 
Therefore, these supersymmetric covariant field strengths will form the 
basic building blocks for constructing higher 
derivative terms invariant under the standard supersymmetry transformations 
(\ref{e2.51}, \ref{e2.52}, \ref{e2.53}). We will look at the effect of such 
terms on the boundary S-matrix coming from the standard supergravity 
action (which we shall compute in section \ref{correl}) 
in section \ref{highder}.

However, as discussed earlier 
there can be other higher derivative terms which can still render the 
full action to be invariant under a modified set of supersymmetry 
transformation laws. Such terms will not necessarily be constructed out of the 
super-covariant field strengths. We will also look at the effect of such 
terms on the standard boundary S-matrix 
towards the end of section \ref{highder}

It is easy to see that 
$G_{MN}$ and $\bar{G}_{MN}$ defined in \refb{EOM} are already covariant 
with respect to the supersymmetry transformations \refb{e2.51}. 
Hence, the fully covariantized 
Rarita-Schwinger field strength $\widetilde{G}_{MN}$ is the same as original 
Rarita-Schwinger field strength $G_{MN}$
\be \label{cov0.1}
\widetilde{G}_{MN}=G_{MN}
\ee
The gauge field strength ${F}_{MN}\equiv 2\p_{[M}A_{N]}$ is not covariant with 
respect to the supersymmetry transformations \refb{e2.51}. 
The covariantization of $F_{MN}$ with respect to supersymmetry is obtained as
\be \label{cov1}
\widetilde{F}_{MN} \equiv 2\partial_{[M}A_{N]}-2\delta^{(Q)}_{\frac{1}{2}\psi_{[M}}A_{N]}
\ee
Using the supersymmetry transformation 
\refb{e2.51}, one finds that $\widetilde{F}_{MN}$ is same as 
$\widehat{F}_{MN}$ defined in \refb{EOM}, i.e
\be \label{cov1.1}
\widetilde{F}_{MN}=\widehat{F}_{MN}=
F_{MN}-\frac{1}{2}\bar{\psi}_{[M}\psi_{N]}
\ee
Now we need to covariantize the
 Riemann tensor. The Riemann tensor $R_{MN}^{~~~~ab}$ (which is the field 
strength associated with $\omega_{M}^{~ab}$) defined in \refb{Ricci} is 
not covariant with respect to the supersymmetry transformations \refb{e2.51} 
and we need to covariantize it. This is obtained as
\be \label{cov2}
\widetilde{\RR}_{MN}^{~~~~ab}= R_{MN}^{~~~~ab}-2\delta^{(Q)}_{\frac{1}{2}\psi_{[M}}\omega_{N]}^{~~ab}, \nonumber \\
\ee
Using the supersymmetry transformations \refb{e2.52} we find
\ben \label{cov2.1}
\widetilde{\RR}_{MN}^{~~~~ab}&=& \RR_{MN}^{~~~~ab}
+\frac{i}{16m}\left[\bar{\psi}_{[M}\gamma_{N]}G^{ab}
-\bar{G}^{ab}\gamma_{[N}\psi_{M]}\right] \nonumber \\
&& \frac{i}{8m}\left[\bar{\psi}_{[M}\gamma^{[a}G_{N]}^{~~b]}
-\bar{G}_{[N}^{~~[b}\gamma^{a]}\psi_{M]}\right] \nonumber \\
\een
The prescriptions \refb{cov1} and \refb{cov2} for covariantization of the 
field strengths have been worked out by looking at the supersymmetry 
transformations considered in section \ref{action}. One can easily check 
that the field strengths reproduced by such prescription are indeed 
covariant with respect to the standard supersymmetry transformations 
(\ref{e2.51}, \ref{e2.52}, \ref{e2.53}). 

\sectiono{Effect of Higher derivative terms} \label{highder}
In this section we will look at the effect of higher derivative 
terms in the action on the boundary S-matrix computed from the 
standard supergravity action\footnote{We shall compute this standard 
boundary S-matrix involving correlation functions of R-symmetry current 
$J(z)$ and supersymmetry current 
$G^{(+)}(z),G^{(-)}(z)$ in section \ref{correl}.}. The most crucial thing to 
note is that by AdS/CFT correspondence 
\cite{Witten:1998qj,Gubser:1998bc}, the action 
evaluated for on-shell configuration of fields, with specified 
boundary conditions, is the partition function for 
evaluating correlators in the boundary, with the boundary values acting as 
sources for relevant correlators\footnote{We will discuss this in great detail 
in section \ref{correl} when we actually evaluate the boundary correlators.}. 
Thus if some terms vanish as a result of using equations of motion, it will 
not contribute to the boundary S-matrix.
 
First let us consider the higher derivative terms appearing in 
$\SSS_{1}$ which is needed 
for the supersymmetrization of the gravitational Chern-Simons term as argued 
in section \ref{action}. This term, as we saw in section \ref{action}, is 
constructed out of 
$\widehat{\RR}_{MN}^{~~ab}$, $\widehat{F}_{MN}$ or $G_{MN}$. So the 
contribution of this term to the equations of motion will be terms containing 
$\widehat{\RR}_{MN}^{~~ab}$, $\widehat{F}_{MN}$, $G_{MN}$ and/or their 
super-covariant derivatives. 
Such terms will necessarily vanish when the original 
equation of motion \refb{EOM}-\refb{eom2} are satisfied. 
Therefore, the solutions 
obtained from $\SSS_{0}$ will continue to hold and $\SSS_{1}$ will vanish 
for such solutions and hence $\SSS_{1}$ will not affect the boundary 
correlators obtained in section \ref{correl}.

There can be other higher derivative terms that can be added to the action 
that are supersymmetric on their own under the standard 
supersymmetry transformations 
(\ref{e2.51}, \ref{e2.52}, \ref{e2.53}). Such terms, as argued in section 
\ref{covfield}, should be constructed 
out of the supersymmetric covariant field strengths 
$\widetilde{\RR}_{MN}^{~~ab}$, $\widehat{F}_{MN}$, $G_{MN}$ and/or 
their supercovariant derivatives. 

Our claim is that these higher derivative terms will change the
 original equations of motion \refb{EOM} by terms which will vanish when the 
original equations of motion \refb{EOM} are satisfied. Therefore, the solution 
obtained for the original 
equations of motion will still solve the 
equations of motion in the presence of 
these higher derivative terms. This, in particular means that 
$\widehat{F}_{MN}=0$ and $G_{MN}=0$ will still continue to hold, 
as a result of 
which the higher derivative terms constructed out of these field strengths 
will vanish and hence will not affect the boundary correlators calculated in 
section \ref{correl}.

Let us analyze our claim in a little detail. The higher derivative terms that 
we could add to the gauge+fermionic sector will be of the form 
\ben \label{hd1}
&& \lambda_1\int d^3 x ~\widehat{F}_{MN}
K_{1}^{MN}[\widehat{F},G,\widetilde{\RR},\DD^{(n)}{(\widetilde{\RR},\widehat{F},G)}]+
\lambda_2\int d^3 x ~\widehat{F}_{MN}K_2^{MN}[{\RR},
D^{(n)}{\RR}] \nonumber \\
&& +\lambda_3 \int d^3 x ~\bar{G}^{M}
\Gamma_{MN}[\widehat{F},G,\widetilde{\RR},
\DD^{(n)}{(\widetilde{\RR},\widehat{F},G)},\gamma]G^{N}
\een
Let us understand the notations used in the above equation. 
($\DD^{(n)}$)$D^{(n)}$ are the $n^{th}$ (super)covariant derivatives. 
$\bar{G}^{M}$ and $G^{M}$ are defined as
\be \label{hd2}
G^{M}=\frac{1}{2}\epsilon^{MNP}G_{NP} \quad 
\bar{G}^{M}=\frac{1}{2}\epsilon^{MNP}\bar{G}_{NP}
\ee
$\Gamma_{MN}$ is a bilinear, constructed out of various supercovariant field 
strengths, their supercovariant derivatives and gamma matrices.

In the first 
integral, $K_{1}^{MN}$ is a function of 
$(\widehat{F},G,\widetilde{\RR},\DD^{(n)}{(\widetilde{\RR}},\widehat{F},G))$. 
Each term of $K_1$ should 
have a factor of $\widehat{F}$, $G$ or their supercovariant derivatives. 
When there are no factors of 
$\widehat{F}$ , $G$ or their supercovariant derivatives involved, need 
special care and hence has been written as 
a separate term in which $K_2$ just depends on 
$({\RR},D^{(n)}{\RR})$\footnote{For supersymmetric 
invariance $K_2$ should be constructed out of the 
supercovariant Riemann tensor $\widetilde{\RR}$ and its 
supercovariant derivatives $\DD^{(n)}\widetilde{\RR}$ but the 
difference between the supercovariant Riemann tensor $\widetilde{\RR}$, 
$\DD^{(n)}\widetilde{\RR}$ and ${\RR}$, $D^{(n)}{\RR}$ are terms proportional 
to $G_{MN}$ and we have included such terms already in $K_{1}^{MN}$.}. 
Since $\widehat{F}_{MN}$ is antisymmetric in $M$ and $N$, 
$K_{2}^{MN}$ should also be antisymmetric in $M$ and $N$. This implies that 
we cannot construct $K_{2}^{MN}$ purely out of ${\RR}_{MN}$ because of the 
symmetric nature of ${\RR}_{MN}$. 
We have to 
involve covariant derivatives of ${\RR}_{MN}$ in each term of $K_2$. 
Therefore, it is obvious that the contribution of the higher derivative terms 
to the equations of motion of $A_M$ and $\psi_{M}$ will have a factor 
of $\widehat{F}_{MN}$, $G_{MN}$, $D_{M}{\RR}_{NP}$ and/or their supercovariant
 derivatives in it. Similarly it can also be argued that the contribution 
to the $\widetilde{\RR}$ equation of motion from the higher derivatives terms 
should have the above factors. 
Just simply factors of ${\RR}_{NP}$ will not 
contribute to any equation of motion. It has to be accompanied by at least one
 of the above factors\footnote{The crucial observation in this analysis 
was that we cannot construct an antisymmetric $K_{2}^{MN}$ purely out of 
${\RR}_{NP}$ because ${\RR}_{NP}$ is symmetric. We have to involve covariant 
derivatives of ${\RR}_{NP}$. The situation would have been drastically 
different if we were in more than three dimension. In that case we could 
construct an antisymmetric second rank tensor by using 
Riemann and Ricci tensors. But in three dimension Riemann tensor is not 
independent and is determined in terms of Ricci tensor and Ricci scalar.}. 
When 
the original equations of motion \refb{EOM}-\refb{eom2} are satisfied, all 
these additional contributions
 to the equations of motion vanish. This justifies our earlier claim that 
the solutions obtained for the original 
equations of motion will still solve the 
equations of motion in the presence of 
these higher derivative terms and that the correlators calculated in section
 \ref{correl} in the gauge+fermionic sector will not be affected by 
these terms. We now use the same supersymmetry argument of 
\cite{David:2007ak} to argue for the non-renormalization of the 
stress tensor correlators. The crux of the argument is that the stress tensor 
correlators are related to the current correlators by supersymmetry. And since 
by our argument the current correlators do not get renormalized, this 
implies that the stress tensor correlators also do not get renormalized.

All the above discussions on the non-renormalization of the boundary S-matrix 
was done in the presence of higher derivative terms which rendered the full 
action to be invariant under the original supersymmetry transformation laws. 
However, as we have said in section \ref{intro} and \ref{covfield}, 
there can be other 
higher derivative terms which can be added that can keep the full action 
invariant under a modified set of supersymmetry transformation laws. As 
argued in \cite{Gupta:2007th} such terms can be removed by field redefinition 
and in terms of the field redefined variables, the action is invariant under 
the original supersymmetry transformation laws. Therefore, such terms will not 
necessarily be constructed out of the super-covariant field strengths.

Before proceeding further, first let us try to understand what sort of terms 
can be removed by redefining $A$ and $\psi$. From \refb{var2} we get
\ben \label{hl1}
\delta \SSS_{0} \over \delta A_M &=& -a_{L}\epsilon^{MNP}\widehat{F}_{NP}
 \nonumber \\
\delta \SSS_{0} \over \delta \psi_M &=& \frac{ia_L}{2}
\epsilon^{MNP}\bar{G}_{NP}=ia_L\bar{G}^M
\nonumber \\
\delta \SSS_{0} \over \delta \bar{\psi}_M &=& \frac{ia_L}{2}
\epsilon^{MNP}{G}_{NP}=ia_L{G}^M
\een
Therefore the terms that can be removed by redefining $A$ and $\psi$ are of 
the form
\be \label{hl2}
\lambda_1\int d^3 x ~\widehat{F}_{MN}
K^{MN} + \lambda_2 \int d^3 x ~\bar{G}^{M}
\Gamma_{MN}G^{N}
\ee
But contrary to \refb{hd1}, $K^{MN}$ and $\Gamma_{MN}$ appearing above are not 
necessarily constructed out of the super-covariant field strengths and 
their super-covariant derivatives. Since such terms can be removed by field 
redefinition, the full action can be invariant in their presence under a 
modified set of supersymmetry transformation laws. 
General coordinate invariance and
 $U(1)$ invariance however suggests that $K^{MN}$ and $\Gamma_{MN}$ should be 
constructed out of $F_{MN}$, $\bar{\psi}_M\psi_{N}$, 
$\bar{\psi}\gamma\psi$ , $\bar{G}_{M}G_{N}$, $\bar{G}\gamma G$ 
$R_{MN}$ and their covariant derivatives. In special cases, $K^{MN}$ and 
$\Gamma_{MN}$ can be functions of the super-covariant field strengths and their
 super-covariant derivatives and \refb{hl2} coincides with \refb{hd1}. 
Such terms 
as we have seen will not modify the equations of motion and will vanish 
on-shell and hence not modify the boundary S-matrix. But a general higher 
derivative term \refb{hl2} can, in principle modify the equations of motion, 
and 
will not vanish on-shell, and hence change the boundary S-matrix. But since 
such terms can be removed by field redefinition, the boundary S-matrix can be 
related to the boundary S-matrix calculated from the standard supergravity 
action by a unitary transformation and such a unitary transformation is exactly
 identity for special class of higher derivative terms considered in \refb{hd1}

\sectiono{Boundary S-matrix} \label{correl}
So long we have been discussing the effects of higher derivative terms 
on the standard boundary S-matrix. Now for the sake of completeness, 
we shall evaluate this standard boundary S-matrix involving 
two and three point correlators of 
the weight ``$1$'' R-symmetry current ($J(z)$) and the 
two weight ``$\frac{3}{2}$'' supersymmetry currents 
($G^{(+)}(z)$ and $G^{(-)}(z)$) in the dual conformal field theory. Here the 
superscript ``$(+)$'' and ``$(-)$'' denote the charges of the corresponding 
operators with respect to the global $U(1)$, which is a symmetry of the theory.

There has been earlier works 
\cite{Rashkov:1999ji,Corley:1998qg,Volovich:1998tj} 
which deals with Rarita-Schwinger fields 
in a general $AdS_{d+1}/CFT_{d}$ correspondence. All the above works 
considered free massless as well as massive Rarita-Schwinger fields without 
coupling to any other fields in the bulk apart from gravity. 
But in a theory of extended supergravity, the Rarita Schwinger field couples 
to gravity as well as gauge field in the bulk. Obtaining the solution to the 
coupled field equations in a general dimension will be a monstrous task. 
However, we will see that in 3 dimensions the 
field equations can be written in a 
form notation and hence solving the coupled equation becomes somewhat simple 
and the coupling to gauge field can be taken care of, in a order by order 
fashion. 

We begin this section by writing the supergravity action in Euclidean space
\ben \label{correl1}
\SSS &=& ia_{L}\int d^3 x ~{\epsilon^{MNP}\left[\frac{1}{2}B_{M}^{~~a}\p_{N}B_{P}^{~~b}
\delta_{ab}+\frac{i}{6}\varepsilon_{abc}B_{M}^{~~a}B_{N}^{~~b}B_{P}^{~~c}\right]} \nonumber \\
&& -ia_{R}\int d^3 x ~{\epsilon^{MNP}\left[\frac{1}{2}B_{M}^{\prime~~a}\p_{N}B_{P}^{\prime~~b}\delta_{ab}+\frac{i}{6}\varepsilon_{abc}B_{M}^{\prime~~a}B_{N}^{\prime~~b}B_{P}^{\prime~~c}\right]} \nonumber \\
&& +i\frac{a_{L}}{2}\int d^3 x ~{\epsilon^{MNP}A_{M}\p_{N}A_{P}} \nonumber \\
&& +\frac{a_L}{4}\int d^3 x ~{\epsilon^{MNP}
\left(\bar{\psi}_{M}\left(\DD_{N}\psi_{P}\right)
-\left(\DD_{N}\bar{\psi}_{M}\right)\psi_{P}\right)}
\een
Where\footnote{Note the change in the definition of $B_{M}^{~~a}$ and 
$B_{M}^{\prime~~a}$ compared to the lorentzian case \refb{e2.3},\refb{e2.4}.}
\ben \label{correl2}
&& B_{M}^{~~a}=\frac{i}{2}\varepsilon^{abc}\omega_{Mbc}-m e_{M}^{~~a}, \nonumber \\
&& B_{M}^{\prime~~a}=\frac{i}{2}\varepsilon^{abc}\omega_{Mbc}+m e_{M}^{~~a}, \nonumber \\
&& \DD_{M}\psi_{N}=\p_{M}\psi_{N}-\frac{1}{2}B_{M}^{~~a}\gamma_{a}\psi_{N}+
\frac{i}{2}A_{M}\psi_{N}, \nonumber \\
&& \DD_{M}\bar{\psi}_{N}=\p_{M}\bar{\psi}_{N}+
\frac{1}{2}B_{M}^{~~a}\bar{\psi}_{N}\gamma_{a}-
\frac{i}{2}A_{M}\bar{\psi}_{N},
\een
The gamma matrices satisfies the algebra:
\ben \label{correl3}
&& \left\{\gamma_{a},\gamma_{b}\right\}= 2\delta_{ab}, \nonumber \\
&& \left[\gamma_{a},\gamma_{b}\right]=-2i\varepsilon_{abc}\gamma_{c},
\een
We will not be interested in obtaining any correlators involving the stress 
tensor. Therefore gravity just enters as a global $AdS_3$ background. The 
metric of Euclidean $AdS_3$ written in Poincare patch coordinate system is
\ben \label{correl5}
ds^2 &=& \frac{1}{m^2(x^0)^2}((dx^0)^2+(dx^1)^2+(dx^2)^2) \nonumber \\
&=& \frac{1}{m^2(x^0)^2}((dx^0)^2+dz d\bar{z}), \quad  \quad z=x^1+ix^2
\een
The gauge field and Rarita Schwinger equations are the same as obtained in 
the Lorentzian case \refb{EOM}. The equations of motion can be written in a 
form notation
\ben \label{correl6}
&& dA=\frac{1}{4}\bar{\psi}\wedge\psi, \nonumber \\
&& d\psi-\frac{1}{2}B^{a}\gamma_{a}\wedge\psi=-\frac{i}{2}A\wedge\psi, 
\nonumber \\
&& d\bar{\psi}+\frac{1}{2}B^{a}\wedge\bar{\psi}\gamma_{a}=
\frac{i}{2}A\wedge\bar{\psi}
\een
Let us now outline the important steps involved in the calculation of the 
correlation function
\begin{enumerate}
\item \label{solve}
Solve the equation \refb{correl6} order by order i.e. 
first solve the leading order equations, 
where we set the RHS appearing in all the equations to zero. Then obtain the 
first order corrections by putting the solution obtained for leading order in 
the RHS and so on. For our purpose we will obtain just the first order 
correction. We will need to impose the following gauge conditions on $\psi$ 
and $\bar{\psi}$ to obtain the solution
\be \label{correl7}
\gamma^M \psi_M=0 \quad \quad \bar{\psi}_M\gamma^M=0
\ee
\item \label{bc}
From the boundary behavior ($x^0 \to 0$), we need to figure out boundary 
values 
of which fields has the appropriate conformal dimension, so as to source the 
corresponding operators in the boundary and then impose boundary conditions 
on those fields.
\item \label{bsol}
Obtain the solutions obtained in \refb{solve} in terms of the 
boundary conditions 
imposed in \refb{bc} but upto terms quadratic in the boundary values.
\item \label{boundy}
We need to add boundary term to the action \refb{correl1} for consistency 
requirements. The reason is the following. 
\noindent
We have imposed boundary conditions on some components of the fields and 
left others to vary. The variation of the action \refb{correl1} will have a 
boundary piece involving the variation of those components of the fields on 
which we have not imposed any boundary conditions and hence will not vanish. 
Therefore we need to add a boundary piece to the action such that its variation
 exactly cancels this contribution.
\item \label{onshell}
Evaluate the bulk action \refb{correl1} as well as boundary action 
obtained in \refb{boundy} as 
a functional of the boundary values but just upto terms cubic in the 
boundary values.
\item \label{cor}
Obtain the correlation function in the boundary using the prescription 
given in \cite{Witten:1998qj,Gubser:1998bc}
\end{enumerate}
Before proceeding with the rest of the section, there is a subtle issue which 
we would like to address. $B_{M}^{~~a}$ and $B_{M}^{\prime~~a}$ defined in 
\refb{correl2} has $\omega_{M}^{~~a}$ in it 
which satisfies the constraint equation 
\refb{spinconnection}. Because of the presence of 
$\bar{\psi}_M\gamma^a \psi_N$ in the right hand side of the equation 
\refb{spinconnection}, 
$\omega_{M}^{~~a}$ and hence $B_{M}^{~~a}$ and $B_{M}^{\prime~~a}$ do not have 
purely gravitational contribution. But the gravitational contribution and 
fermionic contribution to $\omega_{M}^{~~a}$ can be decoupled as 
$\omega_{M}^{~~a}=\omega_{M}^{(e)a}+\omega_{M}^{(f)a}$.

\noindent 
$\omega_{M}^{(e)a}$ satisfies the standard torsionless constraint equation and 
$\omega_{M}^{(f)a}$ contributes to the torsion part of the equation 
\refb{spinconnection}
\ben \label{correl8}
&& de^a+\omega^{(e)ab}\wedge e_b=0, \nonumber \\
&& \omega^{(f)ab}\wedge e_b=-\frac{i}{8m}\bar{\psi}\wedge \gamma^{a} \psi
\een
Clearly $\omega_{M}^{(f)a}$ will be $\OO(\bar{\psi}\gamma \psi)$ and its 
contribution to the equation of motion \refb{correl6} through 
$B_{M}^{~~a}$ and $B_{M}^{\prime~~a}$ will be cubic in $\psi$. Since we would
 like to obtain the solution upto terms quadratic in the fields, we will 
neglect this term and use the standard torsionless spin connection in 
$B_{M}^{~~a}$ and $B_{M}^{\prime~~a}$ for solving the equations of motion 
\refb{correl6}. Since $\omega_{M}^{(f)}$ has a quadratic dependence on the 
fermions, {\it {a Priori}} it seems that, in the evaluation of the action 
for the on-shell configuration of fields as prescribed in step 
\ref{onshell}, as if 
$\omega_{M}^{(f)}$ will give 
a contribution to the on-shell action quadratic in the boundary values of the 
fermions through the dependence of the 
action \refb{correl1} on $B_{M}^{~~a}$ and $B_{M}^{\prime~~a}$ as 
\ben \label{correl9}
&& ia_{L}\int d^3 x ~{\epsilon^{MNP}~\omega_{M}^{(f)a}
\left[\p_{N}B_{Pa}^{(e)}
+\frac{i}{2}\varepsilon_{abc}B_{N}^{(e)b}B_{P}^{(e)c}\right]} \nonumber \\
&& -ia_{R}\int d^3 x ~{\epsilon^{MNP}~\omega_{M}^{(f)a}
\left[\p_{N}B_{Pa}^{(e)\prime}\delta_{ab}
+\frac{i}{2}\varepsilon_{abc}B_{N}^{(e)\prime b}B_{P}^{(e)\prime c}\right]}
\een
Where 
\ben \label{correl10}
B_{N}^{(e) b}=\omega_{N}^{(e)b}-me_{N}^{~~b}, \nonumber \\
B_{N}^{(e)\prime b}=\omega_{N}^{(e)b}+me_{N}^{~~b}
\een
But one can check that for $AdS_3$ background the terms inside the square 
bracket of the above integral \refb{correl9} vanishes and hence will not 
give a quadratic contribution to the evaluation of the on-shell action 
prescribed in step \ref{onshell}. 
However, one has to be careful while obtaining 
quartic and higher contribution. One has to take the effect of the torsion 
in the spin connection into account. But since we are only interested in 
obtaining the on-shell action upto terms cubic in the fields, we will 
use the torsionless spin connection while solving the equation of motion 
\refb{correl6} and while evaluating the action for the on-shell configuration 
of fields we will neglect contribution from the first two terms in the action 
\refb{correl1}.

The stepwise analysis of step \ref{solve} to step 
\ref{onshell} has been given in 
Appendix \ref{analysis}. But we 
will just outline the important results here and obtain the correlators. 
From step \ref{bc} we see that we need to impose the following 
boundary conditions
\footnote{The superscripts in the corresponding boundary values represents 
charge with respect to the $U(1)$ symmetry in the theory.}
\ben \label{correl11}
\lim_{x^0 \to 0}A_{\bar{z}}&=& A_{\bar{z}}^{(0)}(\vec{z}), \nonumber \\
\lim_{x^0 \to 0}(x^0)^{\frac{1}{2}}\psi_{\bar{z}}^{(1)}&=&
\Theta_{\bar{z}}^{(-)}(\vec{z}), \nonumber \\
\lim_{x^0 \to 0}(x^0)^{\frac{1}{2}}\bar{\psi}_{\bar{z}}^{(2)}&=&
\Theta_{\bar{z}}^{(+)}(\vec{z})
\een
From step \ref{boundy} we get the following boundary action
\ben \label{correl12}
&& \SSS_{\texttt{bndy}}=\SSS_{\texttt{bndy}}[\psi,\bar{\psi}]+
\SSS_{\texttt{bndy}}[A] \nonumber \\
&& \SSS_{\texttt{bndy}}[\psi,\bar{\psi}] = -\frac{ia_{L}}{2}
\int~d^2 \vec{z}~\left.
 \left(\bar{\psi}_{z}^{(1)}(x^0, \vec{z})
\psi_{\bar{z}}^{(1)}(x^0, \vec{z})
+\bar{\psi}_{\bar{z}}^{(2)}(x^0, \vec{z})
\psi_{z}^{(2)}(x^0, \vec{z})\right) \right|_{x^0=0} \nonumber \\
&& \SSS_{\texttt{bndy}}[A]=a_{L}\int d^2 \vec{z}~A_{z}(\vec{z},x^0)A_{\bar{z}}(\vec{z},x^0)|_{x^0=0}
\een
Here the superscripts $(1)$ and $(2)$ in the Rarita-Schwinger fields represents
 the spinor components. The measure $d^2{\vec{z}}$ is defined as 
\be \label{correl12.1}
d^2 \vec{z}=dx^1 dx^2
\ee
After obtaining the solution to the equations of motion
 in step \ref{bsol} subject to the boundary conditions \refb{correl11} 
upto terms quadratic 
in the boundary values, 
we proceed to step \ref{onshell} where we evaluate the action 
\refb{correl1} along with the boundary action \refb{correl12} for 
these on-shell 
configuration of fields. We neglect the contribution coming from the first 
two terms in \refb{correl1} as we have argued before. We get
\ben \label{correl13}
\SSS[A_{\bar{z}}^{(0)},\Theta_{\bar{z}}^{(+)},\Theta_{\bar{z}}^{(-)}]
&=&  -\frac{a_L}{\pi}\int d^2 ~\vec{z} ~d^2 \vec{w} ~
\frac{1}{(z-w)^2}~A_{\bar{z}}^{(0)}(\vec{w})~A_{\bar{z}}^{(0)}(\vec{z})
 \nonumber \\
&& -\frac{2ia_L}{\pi}
\int d^2 \vec{z} ~d^2 \vec{w}~\frac{1}{(z-w)^3}~\Theta_{\bar{z}}^{(+)}(\vec{w})~
\Theta_{\bar{z}}^{(-)}(\vec{z}) \nonumber \\
&& -\frac{a_L}{\pi^2}\int d^2 \vec{z} ~d^2 \vec{w} ~d^2 \vec{v}~ 
\frac{1}{(z-w)(z-v)(w-v)^2}~A_{\bar{z}}^{(0)}(\vec{z})~
\Theta_{\bar{z}}^{(+)}(\vec{v})~\Theta_{\bar{z}}^{(-)}(\vec{w}) \nonumber \\
\een
Then according to $AdS/CFT$ conjecture 
\cite{Witten:1998qj,Gubser:1998bc}, we have
\be \label{correl13.1}
\exp({-\SSS(A,\psi,\bar{\psi})})=
\left<\exp\left({\frac{1}{2 \pi}\int_{\p}J(\vec{z})A_{\bar{z}}^{(0)}(\vec{z})
+G^{(+)}(\vec{z})\Theta_{\bar{z}}^{(-)}(\vec{z})+
G^{(-)}(\vec{z})\Theta_{\bar{z}}^{(+)}(\vec{z})}\right)\right>
\ee
This implies the following two and three point correlation functions
\ben \label{correl14}
\left<J(\vec{z_1})J(\vec{z_2})\right>&=&(2\pi)^{2}\left.
\frac{\delta}{\delta A_{\bar{z}}^{(0)}(\vec{z_1})}
\frac{\delta}{\delta A_{\bar{z}}^{(0)}(\vec{z_2})}e^{-\SSS}\right|_{(A_{\bar{z}}^{(0)}(\vec{z}),\Theta_{\bar{z}}^{(+)}(\vec{z}),\Theta_{\bar{z}}^{(-)}(\vec{z}))=0} \nonumber \\
\left<G^{(+)}(\vec{z_1})G^{(-)}(\vec{z_2})\right>&=&(2\pi)^{2}\left.
\frac{\delta}{\delta \Theta_{\bar{z}}^{(-)}(\vec{z_1})}
\frac{\delta}{\delta \Theta_{\bar{z}}^{(+)}(\vec{z_2})}e^{-\SSS}\right|_
{(A_{\bar{z}}^{(0)}(\vec{z}),\Theta_{\bar{z}}^{(+)}(\vec{z}),\Theta_{\bar{z}}^{(-)}(\vec{z}))=0}\nonumber \\
\left<J(\vec{z_1})G^{(+)}(\vec{z_2})G^{(-)}(\vec{z_3})\right>&=&
(2\pi)^{3}\left.
\frac{\delta}{\delta A_{\bar{z}}^{(0)}(\vec{z_1})}
\frac{\delta}{\delta \Theta_{\bar{z}}^{(-)}(\vec{z_2})}
\frac{\delta}{\delta \Theta_{\bar{z}}^{(+)}(\vec{z_3})}e^{-\SSS}\right|_
{(A_{\bar{z}}^{(0)}(\vec{z}),\Theta_{\bar{z}}^{(+)}(\vec{z}),\Theta_{\bar{z}}^{(-)}(\vec{z}))=0} \nonumber \\
\een
Using \refb{correl13} in the above equation \refb{correl14}, we get
\ben \label{correl15}
\left<J(\vec{z_1})J(\vec{z_2})\right>&=& 8a_L \pi \frac{1}{(z_1-z_2)^2}
\nonumber \\
\left<G^{(+)}(\vec{z_1})G^{(-)}(\vec{z_2})\right>&=& 8ia_L \pi 
\frac{1}{(z_1-z_2)^3}  \nonumber \\
\left<J(\vec{z_1})G^{(+)}(\vec{z_2})G^{(-)}(\vec{z_3})\right>&=& 
8a_L \pi\frac{1}{(z_1-z_2)(z_1-z_3)(z_2-z_3)^2}
\een
These are the expected conformal field theory results.
\sectiono{Discussion}
We looked at the $\NN=(2,0)$ supergravity action in the presence of 
gravitational Chern-Simons term in the first order and second order 
formulation. We found some inherent differences between the two formulations. 
In particular, supersymmetry forced us to add a particular type of higher 
derivative term in the second order formulaton and the equations of motion in 
the second order formulation get modified and retain as a subset, the 
solutions to the equations of motion in the first order formulation. 

We then constructed various field strengths covariant with respect to the 
supersymmetry transformations under which the standard supergravity action is 
invariant. These field strengths, as we argued, will form the basic building 
blocks for constructing higher derivative terms which will be invariant under 
the supersymmetry transformations which rendered the standard supergravity 
action to be invariant. 

We then looked at the effect of higher derivative terms on the 
boundary S-matrix which shall arise from the standard supergravity action. 
We saw that there are a special class of higher 
derivative terms constructed out of the super-covariant field strengths 
constructed in section \ref{covfield}. Such terms will not modify the 
equations of motion and will hence vanish on-shell and not modify the 
correlation functions. However, there can be other higher derivative terms, in 
the presence of which, the full action will be invariant under a modified set 
of supersymmetry transformation laws. Such terms can, in principle change the 
equations of motion and will not vanish on-shell as a consequence, and hence 
change the correlation functions. But since such terms can be removed by field 
redefinition, and in terms of the redefined field variables the correlation 
functions are the same as that obtained for the standard action, this implies 
that the correlation functions obtained in the presence of such higher 
derivative terms will be related to the standard correlation functions by a 
unitary transformation. Such a unitary transformation is exactly identity when 
the higher derivative terms are constructed out of the super-covariant field 
strengths and their super-covariant derivatives.

In the end, for the sake of completeness, we computed the standard 
boundary S-matrix describing correlation functions 
of weight ``1'' $U(1)$ current and weight ``$\frac{3}{2}$'' supersymmetry 
current and we found the expected conformal field theory results for the 
two and three point functions.

\acknowledgments
The work of PK was supported by CSIR senior research fellowship and the 
work of BS was supported by DAE senior research fellowship. 
The authors wish to thank Ashoke Sen and Justin David for useful discussions, 
suggestions, comments and careful reading of the manuscript. 
We would specially 
like to thank Justin David for pointing out the references 
\cite{Rashkov:1999ji,Corley:1998qg,Volovich:1998tj} for 
the work 
done involving Rarita-Schwinger fields in a general $AdS_{d+1}/CFT_d$ 
correspondence. We thank Aalok Misra for his support. We 
thank the organizers of the monsoon workshop in string 
theory at TIFR, Mumbai where a part of the work was completed. PK thanks HRI,
 Allahabad for hospitality where most of the work was done.   

\appendix
\section{Boundary S-matrix analysis} \label{analysis}
Here we will give stepwise analysis of the steps outlined in section 
\ref{correl}, 
for obtaining the boundary S-matrix in the gauge+fermionic sector.
\begin{enumerate}
\item
We will present the solution obtained for the equation \refb{correl6} 
till first 
order iteration. For this we need to calculate the vierbeins and the 
torsionless spin connection from the $AdS_3$ metric \refb{correl5} 
and put them in the 
expression for $B_{M}^{~~a}$. We get
\ben \label{a1}
&& B^{\hat{0}}=-\frac{1}{x^{0}}dx^{0}, \nonumber \\
&& B^{\hat{1}}=-\frac{1}{x^{0}}dx^{1}-\frac{i}{x^{0}}dx^{2}=
-\frac{1}{x^{0}}dz, \nonumber \\
&& B^{\hat{2}}=\frac{i}{x^{0}}dx^{1}-\frac{1}{x^{0}}dx^{2}=
\frac{i}{x^{0}}dz
\een
After putting \refb{a1} in the equations of motion \refb{correl6}, 
the leading order equations take the form\footnote{We use 
$\gamma_{\hat{0}}=\sigma_{3}$, $\gamma_{\hat{1}}=\sigma_{1}$, 
$\gamma_{\hat{2}}=-\sigma_{2}$, 
$\gamma_{\hat{z}}\equiv \frac{1}{2}
\left(\gamma_{\hat{1}}-i\gamma_{\hat{2}}\right)$, 
$\gamma_{\hat{\bar{z}}}\equiv \frac{1}{2}
\left(\gamma_{\hat{1}}+i\gamma_{\hat{2}}\right)$}
\ben \label{a2}
&& dA^0=0, \nonumber \\
&& d\psi^0+\frac{1}{2x^0}dx^{0}\wedge\gamma_{\hat{0}}\psi^0+
\frac{1}{x^0}dz\wedge\gamma_{\hat{z}}\psi^0=0, \nonumber \\
&& d\bar{\psi}^0-\frac{1}{2x^0}dx^{0}\wedge\bar{\psi}^0\gamma_{\hat{0}}
-\frac{1}{x^0}dz\wedge\bar{\psi}^0\gamma_{\hat{z}}=0
\een
The solution to the first equation is simple and the solution to the last two 
equations can be obtained after a little algebraic manipulation. We get
\ben \label{a3}
A^0 &=& d\rho, \nonumber \\
\psi^{0(1)} &=& (x^0)^{-\frac{1}{2}}\left(d\eta+\phi dz\right), \nonumber \\
\psi^{0(2)} &=& (x^0)^{\frac{1}{2}}d\phi, \nonumber \\
\bar{\psi}^{0(2)} &=& (x^0)^{-\frac{1}{2}}\left(d\bar{\eta}-\bar{\phi} dz
\right), \nonumber \\
\bar{\psi}^{0(1)} &=& (x^0)^{\frac{1}{2}}d\bar{\phi},
\een
Where, as before the spinor index has been kept in the superscript and in $()$ 
braces and the $0$ to the left of it just denotes the leading order piece. 
Here $\rho$, $\eta$, $\phi$, $\bar{\eta}$, $\bar{\phi}$ are some arbitrary 
functions of ($x^0$, $z$, $\bar{z}$). $\rho$ is an ordinary function 
whereas $\eta$, $\phi$, $\bar{\eta}$, $\bar{\phi}$ are grassman functions.

Now we obtain the first order correction $A^1,\psi^1, \bar{\psi}^1 $ 
to the above equation as
\ben \label{a4}
&& dA^1=\frac{1}{4}\bar{\psi^0}\wedge\psi^0, \nonumber \\
&& d\psi^1+\frac{1}{2x^0}dx^{0}\wedge\gamma_{\hat{0}}\psi^1+
\frac{1}{x^0}dz\wedge\gamma_{\hat{z}}\psi^1=-\frac{i}{2}A^0\wedge\psi^0, 
\nonumber \\
&& d\bar{\psi}^1-\frac{1}{2x^0}dx^{0}\wedge\bar{\psi}^1\gamma_{\hat{0}}
-\frac{1}{x^0}dz\wedge\bar{\psi}^1\gamma_{\hat{z}}= 
\frac{i}{2}A^0\wedge\bar{\psi}^0,
\een
We get
\ben \label{a5}
A^1 &=& \frac{1}{4}\bar{\phi}d\eta+\frac{1}{4}\bar{\eta}d\phi+
\frac{1}{4}\bar{\phi}\phi dz, \nonumber \\
\psi^{1(1)} &=& -\frac{i}{2}(x^0)^{-\frac{1}{2}}\rho (d\eta+\phi dz), 
\nonumber \\
\psi^{1(2)} &=& -\frac{i}{2}(x^0)^{\frac{1}{2}}\rho d\phi, \nonumber \\
\bar{\psi}^{1(2)} &=& \frac{i}{2}(x^0)^{-\frac{1}{2}}\rho 
(d\bar{\eta}-\bar{\phi} dz), 
\nonumber \\
\bar{\psi}^{1(1)} &=& \frac{i}{2}(x^0)^{\frac{1}{2}}\rho d\bar{\phi},
\een
Thus the full solution till first order iteration is
\ben \label{a6}
A &=& d\rho+ \frac{1}{4}\bar{\phi}d\eta+\frac{1}{4}\bar{\eta}d\phi+
\frac{1}{4}\bar{\phi}\phi dz, \nonumber \\
\psi^{(1)}&=&(x^0)^{-\frac{1}{2}}\left(1-\frac{i}{2}\rho\right)(d\eta+\phi dz),
 \nonumber \\
\psi^{(2)}&=&(x^0)^{\frac{1}{2}}\left(1-\frac{i}{2}\rho\right)d\phi, 
\nonumber \\
\bar{\psi}^{(1)}&=&(x^0)^{\frac{1}{2}}\left(1+\frac{i}{2}\rho\right)d\bar{\phi}
, \nonumber \\
\bar{\psi}^{(2)}&=&(x^0)^{-\frac{1}{2}}\left(1+\frac{i}{2}\rho\right)
(d\bar{\eta}-\bar{\phi} dz)
\een
After imposing the gauge condition \refb{correl7}, we get the following 
equations for $\phi$, $\eta$, $\bar{\phi}$ and $\bar{\eta}$
\ben \label{a7}
&& \p_{0}\eta+2x^{0}\p_{\bar{z}}\phi=0 \nonumber \\
&& x^{0}\p_{0}\phi-2\left(\p_{z}\eta+\phi\right)=0 \nonumber \\
&& \p_{0}\bar{\eta}-2x^{0}\p_{\bar{z}}\bar{\phi}=0 \nonumber \\
&& x^{0}\p_{0}\bar{\phi}+2\left(\p_{z}\bar{\eta}-\bar{\phi}\right)=0
\een
The above equations can be solved after decoupling them to obtain a 
second order equation and then using separation of variables to 
separate the $x^0$ equation and $(z, \bar{z})$ equation. The result is
\ben \label{a8}
\eta &=& \frac{1}{2\pi}\int d^2 \vec{p} (x^0 p)^{2}K_2(x^0 p)
e^{i\vec{p}.\vec{z}}\AAA_{\eta}(\vec{p}), \nonumber \\
\phi &=& \frac{1}{2\pi}\int d^2 \vec{p} (x^0 p)K_1(x^0 p)
e^{i\vec{p}.\vec{z}}(-2ip_z)\AAA_{\eta}(\vec{p}), \nonumber \\
\bar{\eta} &=& \frac{1}{2\pi}\int d^2 \vec{p} (x^0 p)^{2}K_2(x^0 p)
e^{i\vec{p}.\vec{z}}\AAA_{\bar{\eta}}(\vec{p}), \nonumber \\
\bar{\phi} &=& \frac{1}{2\pi}\int d^2 \vec{p} (x^0 p)K_1(x^0 p)
e^{i\vec{p}.\vec{z}}(2ip_z)\AAA_{\bar{\eta}}(\vec{p})
\een
Where
\be \label{a8.1}
\vec{p}\equiv (p_z,p_{\bar{z}}) \quad p^2\equiv|\vec{p}|^2\equiv 4 p_z p_{\bar{z}} \quad 
\vec{p}.\vec{z} \equiv p_z z+p_{\bar{z}}\bar{z}
\ee
The measure $d^2 \vec{p}$ is defined as
\be \label{a8.2}
d^2 \vec{p}\equiv dp_1 dp_2
\ee
Where $p_1$ and $p_2$ are related to $p_z$ and $p_{\bar{z}}$ as
\ben \label{a8.3}
&& p_1\equiv\frac{1}{2}\texttt{Re}(p_z), \quad p_2\equiv -\frac{1}{2}\texttt{Im}(p_z) \nonumber \\
&& \texttt{i.e.} p_z=\frac{1}{2}(p_1-ip_2), \quad p_{\bar{z}}=\frac{1}{2}(p_1+ip_2), \nonumber \\
&& \texttt{Hence}~~ p^2 \equiv 4p_z p_{\bar{z}}=p_{1}^{2}+p_{2}^{2} 
\nonumber \\
&& \texttt{And}~~ d^2 \vec{p}\equiv dp_1 dp_2=
\left|\begin{array}{cc}\frac{\p p_1}{\p p_z} &  ~~\frac{\p p_1}{\p p_{\bar{z}}}
\\ \\
\frac{\p p_2}{\p p_z} & ~~\frac{\p p_2}{\p p_{\bar{z}}}\end{array}\right|dp_z dp_{\bar{z}}=-2idp_z dp_{\bar{z}}
\een
Here $K_1$ and $K_2$ are the ``modified Bessel's functions'' 
of order $1$ and $2$
 respectively. They behave asymptotically ($x \to 0$) as
\ben \label{a9}
K_{2}(x) &=& \frac{2}{x^2}\left[1+\OO(x^2)\right], \nonumber \\
K_{1}(x) &=& \frac{1}{x}\left[1+\OO(x^2)\right]
\een
It can be easily checked that the solutions \refb{a8} indeed satisfies the 
equations \refb{a7}. For that, we need to use the following property of 
modified Bessel's functions
\ben \label{a9.1}
K_{2}^{\prime}(x^0 p)&=&-K_{1}(x^0 p)-\frac{2}{x^0 p}K_{2}(x^0 p) \nonumber \\
K_{1}^{\prime}(x^0 p)&=&-K_{2}(x^0 p)+\frac{1}{x^0 p}K_{1}(x^0 p)
\een
\item
We will now figure out the fields on which we should impose boundary 
conditions. Since we are working in $(2,0)$ theory, we should impose boundary 
conditions on the $\bar{z}$ components of the fields. The boundary conditions 
on the gauge field is unambiguous and is simply
\be \label{a10}
\lim_{x^0 \to 0}A_{\bar{z}}(x^0,\vec{z})=A_{\bar{z}}^{(0)}(\vec{z})
\ee
Before trying to figure out what boundary conditions we should impose on 
the Rarita-Schwinger fields, let us outline an important result of 
\cite{Witten:1998qj}. 
According to \cite{Witten:1998qj}, if a p-form field $C$ behave as 
$(x^0)^{-\lambda}C_0$ near the boundary then the operator $\OO$ that couples 
to $C_0$ in the boundary will have conformal dimensions 
$\Delta=d-p+\lambda$. Here $d$ is the dimensions of the boundary. For our case
 $d=2$ and $p=1$ and we want sources for weight $\frac{3}{2}$ supersymmetry 
currents. This suggests $\lambda=\frac{1}{2}$. This behavior near the 
boundary is there for the fields $\psi^{(1)}$ and $\bar{\psi}^{(2)}$ as 
can be seen from \refb{a6}. 
Thus we impose the following boundary conditions on the Rarita-Schwinger 
fields
\ben \label{a11}
\lim_{x^0 \to 0}(x^0)^{\frac{1}{2}}\psi_{\bar{z}}^{(1)}&=&
\Theta_{\bar{z}}^{(-)}(\vec{z}), \nonumber \\
\lim_{x^0 \to 0}(x^0)^{\frac{1}{2}}\bar{\psi}_{\bar{z}}^{(2)}&=&
\Theta_{\bar{z}}^{(+)}(\vec{z})
\een
Following \cite{David:2007ak}, we take $\rho$ appearing in the solutions 
\refb{a6} to be
\ben \label{a12}
&& \rho =\int d^2 \vec{w}~ K(\vec{z},x^0,\vec{w})\BB_{\bar{z}}^{(0)}(\vec{w}),
\nonumber \\
&& K(\vec{z},x^0,\vec{w})=\frac{1}{\pi}\left[\frac{\bar{z}-\bar{w}}{(x^0)^{2}+\left|z-w\right|^{2}}\right]
\een
The function $K(\vec{z},x^0,\vec{w})$ satisfies the following properties on 
the boundary
\ben \label{a13}
&& \lim_{x^0 \to 0}K=\frac{1}{\pi}\frac{1}{z-w}, \nonumber \\
&& \lim_{x^0 \to 0}\p_{z}K=-\frac{1}{\pi}\frac{1}{(z-w)^2}, \nonumber \\
&& \lim_{x^0 \to 0}\p_{\bar{z}}K=\delta^{2}(z-w),
\een
\item
We will use the above properties of $K$ along with the behavior of the 
modified Bessel's functions $K_1$ and $K_2$ near the boundary to fix 
$\BB_{\bar{z}}^{(0)}(\vec{w})$ appearing in \refb{a12} and 
$\AAA_{\eta}(\vec{p})$ and $\AAA_{\bar{\eta}}(\vec{p})$ appearing in 
\refb{a8} in 
terms of the boundary values so that the boundary conditions \refb{a10} and 
\refb{a11} are satisfied. We get
\ben \label{a14}
\BB_{\bar{z}}^{(0)}(\vec{z})&=& A_{\bar{z}}^{(0)}(\vec{z})
~+\frac{1}{4\pi}\int d^2 \vec{w}~\frac{1}{(z-w)^2}
~\Theta_{\bar{z}}^{(+)}(\vec{w})
~\Theta_{\bar{z}}^{(-)}(\vec{z}) \nonumber \\
\AAA_{\eta}(\vec{p})&=&\frac{1}{2\pi}\int ~d^2 \vec{w} 
~\left(\frac{1}{2ip_{\bar{z}}}\right)~e^{-i\vec{p}.\vec{w}}
~\Theta_{\bar{z}}^{(-)}(\vec{w}) \nonumber \\
&& +\frac{i}{4\pi^{2}}\int d^2 \vec{w}~d^2 \vec{v}
~\left(\frac{1}{2ip_{\bar{z}}}\right)~\left(\frac{1}{w-v}\right)
~e^{-i\vec{p}.\vec{w}}
~A_{\bar{z}}^{(0)}(\vec{v})~\Theta_{\bar{z}}^{(-)}(\vec{w}) \nonumber \\
\AAA_{\bar{\eta}}(\vec{p})&=&\frac{1}{2\pi}\int ~d^2 \vec{w} 
~\left(\frac{1}{2ip_{\bar{z}}}\right)~e^{-i\vec{p}.\vec{w}}
~\Theta_{\bar{z}}^{(+)}(\vec{w}) \nonumber \\
&& -\frac{i}{4\pi^{2}}\int d^2 \vec{w}~d^2 \vec{v}
~\left(\frac{1}{2ip_{\bar{z}}}\right)~\left(\frac{1}{w-v}\right)
~e^{-i\vec{p}.\vec{w}}
~A_{\bar{z}}^{(0)}(\vec{v})~\Theta_{\bar{z}}^{(+)}(\vec{w}) \nonumber \\
\een
The above result, then fixes the solutions completely in terms of the boundary 
values. We get\footnote{In order to arrive at the results \refb{a15} we needed 
to do a lot of ``p'' integrals. One of such integral is outlined in the end 
of the section. The others can be similarly obtained}
\ben \label{a15}
A_{z} &=& -\frac{1}{\pi}\int d^2 \vec{w}~\frac{1}{(z-w)^2}~A_{\bar{z}}^{(0)}(\vec{w})
 \nonumber \\
&& -\frac{1}{4\pi^2}\int d^2 \vec{v}~d^2 \vec{w} \frac{1}{(z-w)^2 (w-v)^2}~
\Theta_{\bar{z}}^{(+)}(\vec{v})~\Theta_{\bar{z}}^{(-)}(\vec{w}) \nonumber \\
&& -\frac{1}{2\pi^2}\int d^2 \vec{v}~d^2 \vec{w}~
\frac{1}{(z-w)^3 (z-v)}~\Theta_{\bar{z}}^{(+)}(\vec{v})~
\Theta_{\bar{z}}^{(-)}(\vec{w})+\OO\left((x^0)^2\right)\nonumber \\
A_{\bar{z}} &=& A_{\bar{z}}^{(0)}(\vec{z})+\OO\left((x^0)^2\right)
\nonumber \\
(x^0)^{\frac{1}{2}}\psi^{(1)}_{\bar{z}}&=& \Theta_{\bar{z}}^{(-)}(\vec{z})
 +\OO\left((x^0)^2\right)
\nonumber \\
(x^0)^{\frac{1}{2}}\psi^{(1)}_{z} &=& = \OO\left((x^0)^2\right)\nonumber \\
\nonumber \\
(x^0)^{-\frac{1}{2}}\psi^{(2)}_{z} &=& -\frac{2}{\pi}\int d^2 \vec{w}~ \frac{1}{(z-w)^3}~
\Theta_{\bar{z}}^{(-)}(\vec{w}) \nonumber \\
&& -\frac{i}{\pi^2}\int d^2 \vec{v}~ d^2 \vec{w}~ 
\frac{1}{(z-w)^2 (w-v) (z-v)}~A_{\bar{z}}^{(0)}(\vec{v})~ 
\Theta_{\bar{z}}^{(-)}(\vec{w})+\OO\left((x^0)^2\right) 
\nonumber \\
(x^0)^{-\frac{1}{2}}\psi^{(2)}_{\bar{z}} &=& \OO\left((x^0)^2\right) 
\nonumber \\
 (x^0)^{\frac{1}{2}}\bar{\psi}_{\bar{z}}^{(2)}&=& 
\Theta_{\bar{z}}^{(+)}(\vec{z})+\OO\left((x^0)^2\right)
\nonumber \\
 (x^0)^{\frac{1}{2}}\bar{\psi}_{z}^{(2)}&=& \OO\left((x^0)^2\right) 
\nonumber \\
(x^0)^{-\frac{1}{2}}\bar{\psi}_{z}^{(1)}&=& 
\frac{2}{\pi}\int d^2 \vec{w}~ \frac{1}{(z-w)^3}~ 
\Theta_{\bar{z}}^{(+)}(\vec{w}) \nonumber \\
&& -\frac{i}{\pi^2}\int d^2 \vec{v}~ d^2 \vec{w}~ 
\frac{1}{(z-w)^2 (w-v) (z-v)}~A_{\bar{z}}^{(0)}(\vec{v})~ 
\Theta_{\bar{z}}^{(+)}(\vec{w})+\OO\left((x^0)^2\right) 
\nonumber \\
(x^0)^{-\frac{1}{2}}\bar{\psi}_{\bar{z}}^{(1)}&=&\OO\left((x^0)^2\right)
\een
\item
Now we will obtain the relevant boundary action that we need to add for 
consistency requirements mentioned in section \ref{correl}. Varying the action 
\refb{correl1} we get
\ben \label{a16}
\delta \SSS &=& [0]_{\texttt{on-shell}}+
\frac{ia_L}{2}\lim_{\epsilon \to 0}\int_{\MM_{\epsilon}} ~d^2 \vec{z}~
 \left(\delta{\bar{\psi}_{z}^{(1)}}(\epsilon, \vec{z})
\psi_{\bar{z}}^{(1)}(\epsilon, \vec{z})
+\bar{\psi}_{\bar{z}}^{(2)}(\epsilon, \vec{z})
\delta\psi_{z}^{(2)}(\epsilon, \vec{z})\right) \nonumber \\
&& -a_{L}\lim_{\epsilon \to 0}\int_{\MM_{\epsilon}} d^2 \vec{z}~
\delta A_{z}(\vec{z},\epsilon)A_{\bar{z}}(\vec{z},\epsilon)
\een
Here $[0]_{\texttt{on-shell}}$ denotes the set of terms which vanish on-shell. 
In obtaining the above variation, we have used the fact that we have 
imposed boundary conditions \refb{a10}, \refb{a11} and hence
\be \label{a17}
\lim_{\epsilon \to 0}\delta A_{\bar{z}}(\epsilon,\vec{z})=
\lim_{\epsilon \to 0}\delta \psi_{\bar{z}}^{(1)}(\epsilon,\vec{z})=
\lim_{\epsilon \to 0}\delta \bar{\psi}_{\bar{z}}^{(2)}(\epsilon,\vec{z})=0
\ee
Moreover from \refb{a15}, we observe
\be \label{a18}
\lim_{\epsilon \to 0}\epsilon^{\frac{1}{2}}\psi_{z}^{(1)}(\epsilon,\vec{z})=
\lim_{\epsilon \to 0}\epsilon^{-\frac{1}{2}}\bar{\psi}_{\bar{z}}^{(1)}(\epsilon,\vec{z})=
\lim_{\epsilon \to 0}\epsilon^{\frac{1}{2}}\bar{\psi}_{z}^{(2)}(\epsilon,\vec{z})=
\lim_{\epsilon \to 0}\epsilon^{-\frac{1}{2}}\psi_{\bar{z}}^{(2)}(\epsilon,\vec{z})=0
\ee
The relations \refb{a18} are valid on-shell since these 
are obtained from \refb{a15} which 
are the solutions to the equations of motion. This along with \refb{a17} has 
been used in arriving at \refb{a16}. 
We need to add a boundary action 
$S_{\texttt{bndy}}$ such that
 its variation exactly cancels the integrals in \refb{a16} and we get 
$\delta(S+S_{\texttt{bndy}})=[0]_{\texttt{on-shell}}$. We get
\ben \label{a19}
&& \SSS_{\texttt{bndy}}=\SSS_{\texttt{bndy}}[\psi,\bar{\psi}]+
\SSS_{\texttt{bndy}}[A] \nonumber \\
&& \SSS_{\texttt{bndy}}[\psi,\bar{\psi}] = -\frac{ia_{L}}{2}
\int~d^2 \vec{z}~\left.
 \left(\bar{\psi}_{z}^{(1)}(x^0, \vec{z})
\psi_{\bar{z}}^{(1)}(x^0, \vec{z})
+\bar{\psi}_{\bar{z}}^{(2)}(x^0, \vec{z})
\psi_{z}^{(2)}(x^0, \vec{z})\right) \right|_{x^0=0} \nonumber \\
&& \SSS_{\texttt{bndy}}[A]=a_{L}\int d^2 \vec{z}~A_{z}(\vec{z},x^0)A_{\bar{z}}(\vec{z},x^0)|_{x^0=0}
\een
\item
Now we will put the solutions obtained in \refb{a15} in the bulk as well as 
boundary action and obtain the on-shell action as a functional of the boundary 
values. The contribution from the boundary action is straightforward to obtain.
 We get
\ben \label{a20}
\SSS_{\texttt{bndy}}[\psi,\bar{\psi}]&=& -\frac{2ia_L}{\pi}
\int d^2 \vec{z} ~d^2 \vec{w}~\frac{1}{(z-w)^3}
~\Theta_{\bar{z}}^{(+)}(\vec{w})~
\Theta_{\bar{z}}^{(-)}(\vec{z}) \nonumber \\
&& -\frac{a_L}{\pi^2}\int d^2 \vec{z} ~d^2 \vec{w} ~d^2 \vec{v}~ 
\frac{1}{(z-w)(z-v)(w-v)^2}~A_{\bar{z}}^{(0)}(\vec{z})~
\Theta_{\bar{z}}^{(+)}(\vec{v})~\Theta_{\bar{z}}^{(-)}(\vec{w}) \nonumber \\
\nonumber \\
\SSS_{\texttt{bndy}}[A]&=& -\frac{a_L}{\pi}\int d^2 ~\vec{z} ~d^2 \vec{w} ~
\frac{1}{(z-w)^2}~A_{\bar{z}}^{(0)}(\vec{w})~A_{\bar{z}}^{(0)}(\vec{z})
 \nonumber \\
&& -\frac{a_L}{4\pi^2}\int d^2 \vec{z} ~d^2 \vec{w} ~d^2 \vec{v}~
\frac{1}{(z-w)^2 (w-v)^2}~A_{\bar{z}}^{(0)}(\vec{z})~
\Theta_{\bar{z}}^{(+)}(\vec{v})~\Theta_{\bar{z}}^{(-)}(\vec{w})
 \nonumber \\
&& -\frac{a_L}{2\pi^2}\int d^2 \vec{z} ~d^2 \vec{w} ~d^2 \vec{v}~
\frac{1}{(z-w)^3 (z-v)}~A_{\bar{z}}^{(0)}(\vec{z})~
\Theta_{\bar{z}}^{(+)}(\vec{v})~\Theta_{\bar{z}}^{(-)}(\vec{w}) \nonumber \\
\een
In order to obtain the contribution from the bulk action \refb{correl1}, we 
recall from section \ref{correl} that the first two terms in the action will 
not contribute towards the evaluation of two and three point functions. The 
last term vanish by equations of motion. So the only term that contributes is
\ben \label{a21}
\SSS_{\texttt{bulk}}&=&
i\frac{a_{L}}{2}\int d^3 x ~{\epsilon^{MNP}A_{M}\p_{N}A_{P}} \nonumber \\
&=& i\frac{a_{L}}{2}\int d^3 x ~{\epsilon^{MNP}A_{M}^{0}\p_{N}A_{P}^{0}}
+i\frac{a_{L}}{2}\int d^3 x ~{\epsilon^{MNP}A_{M}^{0}\p_{N}A_{P}^{1}}
\nonumber \\
&& i\frac{a_{L}}{2}\int d^3 x ~{\epsilon^{MNP}A_{M}^{1}\p_{N}A_{P}^{0}}
+i\frac{a_{L}}{2}\int d^3 x ~{\epsilon^{MNP}A_{M}^{1}\p_{N}A_{P}^{1}} 
\nonumber \\
\een
In the above equation, 
we have broken the full solution to a sum of the leading 
solution and the first order correction. Since $A_{M}^{0}=\p_M \rho$, this 
implies $\epsilon^{MNP}\p_N A_{P}^{0}=0$. The last term in \refb{a21} 
gives a quartic 
contribution and hence can be neglected. Thus
\ben \label{a22}
S_{\texttt{bulk}}[A,\psi,\bar{\psi}]&=& i\frac{a_{L}}{2}\int d^3 x~ 
\epsilon^{MNP}~A_{M}^0 ~\p_{N}A_{P}^1 \nonumber \\
&=& i\frac{a_{L}}{2}\int d^3 x~ 
\epsilon^{MNP}~\p_{N}\left(A_{M}^0 ~A_{P}^1\right) \nonumber \\
&=& a_L\int d^2 \vec{z}\left. ~\left(A_{z}^0(\vec{z}) ~A_{\bar{z}}^1(\vec{z})-A_{\bar{z}}^0(\vec{z}) ~A_{z}^1(\vec{z})\right)\right|_{x^0=0} \nonumber \\
\een
In order to evaluate the above contribution from the bulk action we need to 
know the leading order solution and first order correction separately for the 
gauge field $A_M$.
\ben \label{a23}
A^{0}_{\bar{z}}&=& \p_{\bar{z}}\rho \nonumber \\
&=& A_{\bar{z}}^{(0)}(\vec{z})+\frac{1}{4\pi}\int d^2 \vec{w}
 \frac{1}{(z-w)^2}~\Theta_{\bar{z}}^{(+)}(\vec{w})~
\Theta_{\bar{z}}^{(-)}(\vec{z}), 
\nonumber \\
A^{0}_{z}&=& \p_{z}\rho \nonumber \\
&=& -\frac{1}{\pi}\int d^2 \vec{w}~\frac{1}{(z-w)^2}~A_{\bar{z}}^{(0)}(\vec{w})
 \nonumber \\
&& -\frac{1}{4\pi^2}\int d^2 \vec{v}~d^2 \vec{w} \frac{1}{(z-w)^2 (w-v)^2}~
\Theta_{\bar{z}}^{(+)}(\vec{v})~\Theta_{\bar{z}}^{(-)}(\vec{w})
 \nonumber \\
A^{1}_{z} &=& \frac{1}{4}\bar{\phi}\left(\p_z \eta +\phi \right)+ 
\frac{1}{4}\bar{\eta}\p_{z}\phi \nonumber \\
&=& -\frac{1}{2\pi^2}\int d^2 \vec{v}~d^2 \vec{w}~
\frac{1}{(z-w)^3 (z-v)}~\Theta_{\bar{z}}^{(+)}(\vec{v})~
\Theta_{\bar{z}}^{(-)}(\vec{w}) \nonumber \\
 \nonumber \\
A^{1}_{\bar{z}} &=& \frac{1}{4}\bar{\phi}\p_{\bar{z}}\eta+
\frac{1}{4}\bar{\eta}\p_{\bar{z}}\phi \nonumber \\
&=& -\frac{1}{4\pi}\int d^2 \vec{w}~
\frac{1}{(z-w)^2}~\Theta_{\bar{z}}^{(+)}(\vec{w})~
\Theta_{\bar{z}}^{(-)}(\vec{z}) \nonumber \\
\een
Putting these solutions in \refb{a22}, we get the contribution from the 
bulk action upto terms cubic in the boundary values as 
\ben \label{a24}
\SSS_{\texttt{bulk}}[A,\psi,\bar{\psi}] &=& a_L\int d^2 \vec{z}~\left(A_{z}^0(\vec{z}) ~A_{\bar{z}}^1(\vec{z})-A_{\bar{z}}^0(\vec{z}) ~A_{z}^1(\vec{z})\right)
\nonumber \\
&=& \frac{a_L}{4\pi^2}
\int d^2 \vec{z}d^2 \vec{v} d^{2}\vec{w}~\frac{1}{(z-w)^2 (w-v)^2}~
A_{\bar{z}}^{(0)}(\vec{z})~
\Theta_{\bar{z}}^{(+)}(\vec{v})~\Theta_{\bar{z}}^{(-)}(\vec{w}) \nonumber \\
&& +\frac{a_L}{2\pi^2}
\int d^2 \vec{z}d^2 \vec{v} d^{2}\vec{w}~\frac{1}{(z-v) (z-w)^3}~
A_{\bar{z}}^{(0)}(\vec{z})~
\Theta_{\bar{z}}^{(+)}(\vec{v})~\Theta_{\bar{z}}^{(-)}(\vec{w})\nonumber \\
\een
Combining this with the boundary contributions obtained in \refb{a20}, we get
\ben \label{a25}
\SSS[A,\psi,\bar{\psi}]&=& \SSS_{\texttt{bulk}}[A,\psi,\bar{\psi}]+
\SSS_{\texttt{bndy}}[\psi,\bar{\psi}]+S_{\texttt{bndy}}[A] \nonumber \\
&=&  -\frac{a_L}{\pi}\int d^2 ~\vec{z} ~d^2 \vec{w} ~
\frac{1}{(z-w)^2}~A_{\bar{z}}^{(0)}(\vec{w})~A_{\bar{z}}^{(0)}(\vec{z})
 \nonumber \\
&& -\frac{2ia_L}{\pi}
\int d^2 \vec{z} ~d^2 \vec{w}~\frac{1}{(z-w)^3}~\Theta_{\bar{z}}^{(+)}(\vec{w})~
\Theta_{\bar{z}}^{(-)}(\vec{z}) \nonumber \\
&& -\frac{a_L}{\pi^2}\int d^2 \vec{z} ~d^2 \vec{w} ~d^2 \vec{v}~ 
\frac{1}{(z-w)(z-v)(w-v)^2}~A_{\bar{z}}^{(0)}(\vec{z})~
\Theta_{\bar{z}}^{(+)}(\vec{v})~\Theta_{\bar{z}}^{(-)}(\vec{w}) \nonumber \\
\een
\end{enumerate}
Obtaining the correlation functions from here on is straightforward and has 
been obtained in section \ref{correl}.

In order to arrive at the solutions of the fields \refb{a15} in terms of the 
boundary values, 
we had to carry out a number of non-trivial ``$p$'' integrals. 
We conclude this section by outlining one of such integrals. The other 
integrals can be similarly worked out. One of such integrals is
\be \label{a26}
I_{2}(\vec{z}-\vec{w})\equiv -\frac{i}{4\pi^{2}}\int~d^2 \vec{p}~\left(\frac{p_{z}^{2}}{p_{\bar{z}}}\right)e^{i\vec{p}.(\vec{z}-\vec{w})}
\ee
In order to evaluate the above integral, we define
\ben \label{a29}
&& \Upsilon \equiv p_{z}(z-w) \quad \quad 
\bar{\Upsilon} \equiv p_{\bar{z}}(\bar{z}-\bar{w}) \nonumber \\
\Rightarrow && d^2 \vec{p} \equiv -2i dp_z dp_{\bar{z}}=
\frac{-2i}{(z-w)(\bar{z}-\bar{w})}d\Upsilon d\bar{\Upsilon} \nonumber \\
&& ~~~~~~~~~~~~~~~~~~~~~~~\equiv \frac{1}{(z-w)(\bar{z}-\bar{w})}d^2 \vec{\Upsilon}
\een
$d^2 \vec{\Upsilon}$, similarly to $d^2 \vec{p}$ in \refb{a8.2} and 
\refb{a8.3}, can be 
expressed in terms of its real and imaginary parts
\ben \label{a30}
d^2 \vec{\Upsilon}\equiv d\Upsilon_{1}d\Upsilon_{2} \nonumber \\
\texttt{where} ~~\Upsilon_{1}\equiv \frac{1}{2}\texttt{Re}(\Upsilon), \quad 
\Upsilon_{2}\equiv -\frac{1}{2}\texttt{Im}(\Upsilon), \nonumber \\
\een
With all the above definitions, the integral \refb{a26} 
takes the following form
\ben \label{a31}
I_{2}(\vec{z}-\vec{w})&=& -\frac{i}{4\pi^{2}}\frac{1}{(z-w)^3}\int~d^2 \vec{\Upsilon}~\left(\frac{\Upsilon^{2}}{\bar{\Upsilon}}\right)e^{i(\Upsilon+\bar{\Upsilon})} \nonumber \\
&\equiv& \frac{c_2}{(z-w)^3}
\een
The last line of the above equation defines $c_2$ as
\ben \label{a32}
c_2 \equiv -\frac{i}{4\pi^{2}}\int~d^2 \vec{\Upsilon}~\left(\frac{\Upsilon^{2}}{\bar{\Upsilon}}\right)e^{i(\Upsilon+\bar{\Upsilon})}
\een
Thus the ``$p$'' integral \refb{a26} now boils down to doing the 
``$\Upsilon$'' integral \refb{a32}. In order to do the integral \refb{a32} we 
define
\ben \label{a33}
&& \Upsilon_{1}=R\cos\theta, \quad \Upsilon_{2}=R\sin\theta \nonumber \\
\Rightarrow && d^2 \vec{\Upsilon}=R ~dR ~d\theta, \quad 
\Upsilon=\frac{1}{2}Re^{-i\theta}, \quad
\bar{\Upsilon}=\frac{1}{2}Re^{i\theta}, \quad 
\Upsilon+\bar{\Upsilon}=R\cos\theta \nonumber \\
\een
Putting this definition in \refb{a32}, we get
\ben \label{a34}
c_2 &=& -\frac{i}{4\pi^{2}}\frac{1}{2}\int_{0}^{\infty}~R^2~dR\int_{0}^{2\pi}~e^{i(Rcos\theta-3\theta)}d\theta, \nonumber \\
&=& -\frac{1}{4\pi}\int_{0}^{\infty} R^2 J_{3}(R)~dR = -\frac{2}{\pi}
\een
In the last step, we have used the integral representation of Bessel's function
\be \label{a35}
J_{n}(x)=\frac{i^{-n}}{2\pi}\int_{0}^{2\pi} e^{i(x\cos\theta+n\theta)}~d\theta
\ee
And
\be \label{a36}
\int_{0}^{\infty}J_{n}(R)~dR =1 \quad \int_{0}^{\infty}RJ_{n-1}(R)~dR = n-1 \quad \int_{0}^{\infty}R^{2}J_{n-1}(R)~dR=n(n-2)
\ee
Therefore the integral \refb{a26} takes the form
\be \label{a37}
I_{2}(\vec{z}-\vec{w})=-\frac{2}{\pi(z-w)^3}
\ee
 \bibliography{corel}
\bibliographystyle{JHEP}

\end{document}